\documentclass[prc,twocolumn,amsmath,amssymb,superscriptaddress,noffotinbib,preprintnumbers]{revtex4-1}
\usepackage{url,color,colordvi,epsfig,pstricks}
\usepackage[colorlinks=true, citecolor=blue, filecolor=blue, linkcolor=blue, urlcolor=blue, pdftex]{hyperref}
\usepackage{graphicx,bm,dcolumn}
\usepackage{hyperref,slashed,lineno}
\usepackage{physics}
\usepackage{inputenc}

\usepackage{soul}

\hypersetup{
    colorlinks=true, 
    linktoc=all, 
    linkcolor=blue, 
    citecolor=blue}

\newcommand{\np}{\mathbf{p}}

\usepackage{xspace}
\newcommand{\nuwro}{\textsc{NuWro}\xspace}
\newcommand{\incl}{\textsc{INCL}\xspace}
\newcommand{\abla}{\textsc{ABLA}\xspace}

\newcommand{\diff}{\mathrm{d}}						 								

\begin{document}

\title{Final-state interactions in neutrino-induced proton knockout from argon in MicroBooNE}
\author{A.~Nikolakopoulos}\email{anikolak@fnal.gov}
\affiliation{Theoretical Physics Department, Fermilab, Batavia IL 60510, USA}
\author{A.~Ershova}
\affiliation{Laboratoire Leprince-Ringuet, Ecole polytechnique, IN2P3-CNRS, Palaiseau, France}
\author{R.~Gonz\'alez-Jim\'enez}
\affiliation{Grupo de F\'isica Nuclear, Departamento de Estructura de la Materia, F\'isica T\'ermica y Electr\'onica, Facultad de Ciencias F\'isicas, Universidad Complutense de Madrid and IPARCOS, CEI Moncloa, Madrid 28040, Spain}
\author{J.~Isaacson}
\affiliation{Theoretical Physics Department, Fermilab, Batavia IL 60510, USA}
\author{A.~M.~Kelly}
\affiliation{Theoretical Physics Department, Fermilab, Batavia IL 60510, USA}
\author{K.~Niewczas}
\affiliation{Department of Physics and Astronomy, Ghent University, Proeftuinstraat 86, 9000 Gent, Belgium}
\author{N.~Rocco}
\affiliation{Theoretical Physics Department, Fermilab, Batavia IL 60510, USA}
\author{F.~S\'anchez}
\affiliation{Univserity of Geneva, Section de Physique, DPNC, 1205 Geneva Switzerland}

\begin{abstract}
    Neutrino event generators make use of intranuclear cascade models (INCs), to predict the kinematics of hadron production in neutrino-nucleus interactions. We perform a consistent comparison of different INCs, by using the same set of events as input to the NEUT, \nuwro, \textsc{Achilles} and \incl INCs. The inputs correspond to calculations of the fully differential single-proton knockout cross section, either in the distorted-wave impulse approximation (DWIA) or plane-wave impulse approximation (PWIA), both including realistic nuclear hole spectral functions. We compare the INC results to DWIA calculations with an optical potential, used extensively in the analysis of (e,e'p) experiments. We point out a systematic discrepancy between both approaches. We apply the INC results to recent MicroBooNE data. We assess the influence of the choice of spectral function, finding that large variations in realistic spectral functions are indistinguishable with present data. 
    The data is underpredicted, with strength missing in the region where two-nucleon knockout and resonance production contribute. However, the data is underpredicted also in regions of low transverse missing momentum, where one-nucleon knockout dominates. The inclusion of the interference with two-body currents could lead to additional strength in this region.
\end{abstract}

\maketitle


\section{Introduction}
Many modern accelerator-based neutrino experiments, such as the Short Baseline Neutrino (SBN) program (SBND, MicroBooNE, and ICARUS) at Fermilab, and the future Deep Underground Neutrino Experiment (DUNE) use Liquid Argon Time Projection Chambers (LArTPCs) to measure the hadrons produced in neutrino-argon interactions.

The simplest neutrino-nucleus interaction involving the emission of final-state hadrons is arguably charged-current single nucleon knockout, where a muon and proton are detected in coincidence, and the residual system is left in a low-energy excited state.
In neutrino experiments however, such a process cannot be uniquely isolated, because the energy transferred to the nuclear system is unknown on an event-by-event basis.
Indeed, in typical nucleon knockout experiments (e.g., electron-induced proton knockout) the total incoming energy is known. This allows one to select kinematics to restrict the excitation energy of the unobserved system, thereby restricting the signal to true single proton knockout.
In neutrino experiments, measurements are instead one-muon-one-proton events, with a partially characterized additional hadron system.
Several such measurements have been performed by the MicroBooNE collaboration in recent years~\cite{MicroBooNE:ddifshort2023, MicroBoonE:ddiflong2023, MicroBooNE:2024tmp, MicroBooNE:2024xod, MicroBooNE:2023krv}.
It is clear that to describe such data, one needs to take into account the whole spectrum of possible final states for the hadron system. Unfortunately, a suitable microscopic description of such a signal is presently unavailable.
At present, neutrino event generators, which serve as a theoretical input into experimental analyses, describe the process as happening in two steps. First an initial interaction occurs with one or two nucleons, thereby populating a limited (although infinite) number of final states. The energy and momentum of hadrons produced in this initial step are then redistributed through strong secondary interactions with the rest of the nucleus.
For this second step use is made of intranuclear cascade models (INCs) or kinetic transport theory as in GiBUU~\cite{Buss12}. 
The two-step approach can be motivated by the fact that the inclusive cross section can be described by an incoherent sum of different interaction mechanisms, typically, quasielastic scattering (QE), two-nucleon knockout and inelastic interactions with single nucleons, and that in such a description a limited number of final-states needs to be taken into account for the residual system~\cite{Benhar08,Amaro20}. 
The secondary interactions then redistribute this strength over final-state configurations in a unitary way, i.e. without altering the inclusive cross section.  
We will refer to this second step as \emph{inelastic} final-state interactions (FSI), as it requires the exchange of energy between the hadron(s) produced in the primary interaction and the residual system.

Comparisons of several event generators and INCs to recent MicroBooNE data have been presented in Refs.~\cite{MicroBooNE:ddifshort2023, MicroBoonE:ddiflong2023, MicroBooNE:2024tmp}. These studies, and data taken under similar experimental conditions with electrons~\cite{CLAS:2021neh}, show that no single approach provides a satisfying reproduction of all the data. 
The input used for INCs in neutrino generators varies widely, which makes it impossible to disentangle the effect of inelastic FSI interactions from the description of the initial interaction in these comparisons. Moreover, some approaches that provide a reasonable description of the inclusive cross section, such as the SuSAv2 approach~\cite{SuSav2, SuSAMEC}, do not provide information on the produced final-state hadrons at all, leading to approximations being made to determine the outgoing nucleon kinematics~\cite{Dolan19, Nikolakopoulos:2023pdw}. Others, such as those based on the (local) Fermi gas~\cite{Nieves11, Martini16, Martini:2022ebk}, while providing a reasonable description of some inclusive and semi-inclusive~\cite{Bourguille:2020bvw} observables are incompatible with our knowledge of the shell structure of nuclei~\cite{VOrden2019, BENHAR:RevModPhyseep}.

In this work, we bridge the gap between the description of the inclusive and exclusive cross sections in the INC picture. 
We disentangle the effect of inelastic FSI from the initial interaction, by using consistent inputs for different INCs. 
We present results for the INCs implemented in \nuwro~\cite{NuWroFSI,Niewczas:2019fro}, NEUT~\cite{Hayato:NEUT, Hayato:2021heg}, \textsc{Achilles}~\cite{Isaacson:2020wlx, Isaacson:2022cwh}, and the Li\`ege INC (\incl)~\cite{Boudard:2012wc, Rodriguez-Sanchez:2017odk}.
For the inputs, we use realistic spectral functions, including partial occupations of mean-field states, and a contribution from short-range correlations~\cite{BENHAR:RevModPhyseep}.
We use calculations in both the (relativistic) plane-wave impulse approximation (R)PWIA, neglecting Pauli blocking and all FSI effects, and the relativistic distorted-wave impulse approximation (RDWIA). The latter use the energy-dependent relativistic mean-field potential of Refs.~\cite{Gonzalez-Jimenez19, Gonzalez-Jimenez20}, to include Pauli blocking and the \emph{elastic} FSI necessary to describe the inclusive cross section. What we dub elastic FSI for the purpose of this paper, is distinct from the inelastic FSI included through the INC. The dispersion relation of the outgoing nucleon is altered in the medium, but no exchange of energy (other than recoil energy) with the residual system happens. Unlike the INC, the elastic FSI does change the inclusive cross section with respect to the RPWIA.  
The difference with the approach of Refs.~\cite{Ankowski:betterE} to include essentially the same effect, is that the latter is applied to the cross section after integration over nucleon kinematics. In the RDWIA, the fully differential nucleon knockout cross section can be computed with the necessary elastic FSI.

As in Ref.~\cite{Nikolakopoulos:2022qkq}, we compare the results of these INCs for direct nucleon knockout to optical potential calculations. This benchmark calculation corresponds to the approach used in analyses of $(e,e' p)$, in particular the recent analyses for argon and titanium targets~\cite{PhysRevD.105.112002, PhysRevD.107.012005}.
We provide a detailed comparison of the different INCs for several kinematic variables, where the results are averaged over the MicroBooNE flux.
We show that the main differences between the INCs are found at low proton-energy, and from the treatment of the effect of short-range correlations on the mean-free path~\cite{PhysRevC.45.791}.
We discuss the significant amount of nuclear clusters produced in the \incl cascade, and in particular point out that the production of deuteron should be measurable in LArTPCs as previously reported for calculations for the T2K experiment~\cite{Ershova:2022jah, Ershova:2023dbv}.

This work expands on the results of Refs.~\cite{Franco-Patino:2023msk, Franco-Patino:2022tvv, Gonzalez-Jimenez:2021ohu}, by including the INC to model inelastic FSI explicitly through rescattering, which is shown to be non-negligible for all observables studied.
The recent results of Ref.~\cite{Banerjee:2023hub} obtained with \nuwro are similar, using the spectral function of Refs.~\cite{PhysRevD.105.112002, PhysRevD.107.012005} in the plane-wave impulse approximation.
The latter include the contribution of meson-exchange currents and resonance production, but lack the effect of elastic FSI.

The approach of the present work, combining the fully differential RDWIA calculation suitable for the inclusive cross section with realistic spectral functions and the INC, 
is clearly a step forward in the modeling of neutrino-induced nucleon knockout.
We hence compare the results to the recent MicroBooNE data~\cite{MicroBooNE:ddifshort2023, MicroBoonE:ddiflong2023}. 
We find a 10 $\%$ effect of elastic FSI on observables, compared to the RPWIA calculations,
and a significant model dependence for the INC.
The data is underpredicted overall, as expected, because of the lack of additional interaction mechanisms. 
Interestingly, the data is also underpredicted at kinematics where direct nucleon knockout dominates, which might be, at least partially, resolved by including the interference with two-body currents~\cite{Franco-Munoz:2022jcl, Lovato:2023khk}.
This interpretation is speculative however, as there is a model-dependent contribution of rescattered events for these kinematics, and the data are presented in terms of reconstructed variables, which include a smearing of true kinematics.

In the following, section~\ref{secCS}, the methodology is described, we discuss modeling of the inclusive cross section, the exclusive cross section, the spectral function and the INC approach. Details for the calculations are given in appendix.~\ref{app:hadronT}. Results are shown in section~\ref{sec:results}, we show the influence of the choice of spectral function in~\ref{sec:SFcomp}, the comparison between INCs in~\ref{sec:INC_ROP}, and the comparison to MicroBooNE data in~\ref{sec:Micro_comp}.  Our final remarks are presented in section~\ref{conclusions}.

\section{Methodology}{\label{secCS}}
In this work, we describe semi-inclusive one-nucleon knockout on argon for the kinematics of the MicroBooNE experiment~\cite{MicroBoonE:ddiflong2023, MicroBooNE:ddifshort2023}.
Our approach uses and compares the results of several intranuclear cascade models (INCs) (NEUT~\cite{Hayato:NEUT, Hayato:2021heg}, \nuwro~\cite{WRONG,NuWro:FSI}, \textsc{Achilles}~\cite{Isaacson:2022cwh, Isaacson:2020wlx} and \incl~\cite{Boudard:2012wc, Rodriguez-Sanchez:2017odk}) to model inelastic final-state interactions.
The input to the INCs are obtained with calculations of the full five-fold differential cross section, including realistic nuclear spectral functions, and the effect of nucleon distortion in a real potential.
The output of the classical INC approach is compared to quantum mechanical calculations of exclusive one-nucleon knockout which use an identical description of the nuclear initial-state~\cite{Nikolakopoulos:2022qkq}. The latter are obtained in the relativistic distorted-wave impulse approximation with an optical potential~\cite{Cooper93}.

These different ingredients are described in the following.
The general formalism and kinematics for flux-averaged semi-inclusive cross sections are summarized in Sec.~\ref{sec:kinandcross}.
Sec.~\ref{sec:Exclusive} is devoted to the optical potential and spectral functions. Finally Secs.~\ref{sec:INCs} and~\ref{sec:inputinclusive} discuss the INC approach and the calculations used as input to the INC respectively.

\subsection{Kinematics and cross section}
\label{sec:kinandcross}
We consider the semi-inclusive proton knockout interaction $A(\nu_\mu, \mu^- p)X$. For this process, the particle four-momenta are written as
\begin{equation}
k_{\nu} + k_A = k_N + k_{\mu} + k_X,
\end{equation}
where $k_A$ is the initial nucleus, $k_N$ the knocked out proton and $k_X$ is the residual hadron system which remains (partly) unobserved.
The four-momentum transferred to the nuclear system is denoted
\begin{equation}
Q = k_\nu - k_\mu,
\end{equation}
and the squared four-momentum transfer 
\begin{equation}
Q^2 \equiv - Q \cdot Q = q^2 - \omega^2,
\end{equation}
is defined as positive as usual. The energy and momentum transfer to the nucleus are denoted $\omega = E_\nu - E_\mu$ and $q = \lvert \mathbf{p}_\nu - \mathbf{p}_\mu \rvert$, respectively. We denote three-momenta in bold, and their magnitude as $p$.

In exclusive measurements, the kinematics of all particles are uniquely specified.
Energy and momentum transfer, along with an outgoing proton momentum $\mathbf{p}_N$, completely specify the invariant mass and energy of the residual system by
\begin{equation}
\omega + M_A = \sqrt{\mathbf{p}_N^2 + M_N} + \sqrt{\mathbf{p}_X^2 + M_X^2},
\end{equation}
and with the missing momentum defined as
\begin{equation}
\mathbf{p}_m \equiv -\mathbf{p}_X = \mathbf{p}_N - \mathbf{q}.
\end{equation}
The missing energy is 
\begin{equation}
E_m \equiv M_X + M_N - M_A = E_\nu - E_\mu - T_N - T_X,
\end{equation}
where $T_N, T_X$ are the kinetic energy of the nucleon and residual system respectively.

The setup in accelerator-based neutrino experiments greatly complicates the description of the final-state hadron system compared to the exclusive case, where e.g. production of a single excited final-state with missing energy below two-nucleon knockout threshold is considered.
Neutrino experiments are instead exposed to a broad flux of incoming neutrinos, and the incoming energy is not known on an event-by-event basis.
As such, we consider semi-inclusive cross sections~\cite{VOrden2019, Gonzalez-Jimenez:2021ohu}.
The probability for a $A(\nu_\mu, \mu^- p)X$ event is proportional to the flux-averaged cross section
\begin{align}
\label{eq:sigma_fluxfolded}
& \left\langle \frac{\diff \sigma }{\diff p_\mu \diff \Omega_\mu \diff p_N \diff \Omega_N}\right\rangle = \frac{p_\mu^2 p_N^2 M_N}{(2\pi)^5 E_\mu E_N} \\
&\times   \int \diff E_\nu \frac{\Phi(E_\nu)}{E_\nu} \delta(\omega + M_A - E_N - E_X) \sum_{f} \lvert \mathcal{M}_{fi} \rvert^2,
\end{align}
with $\Phi(E_\nu)$ denoting the neutrino flux.
The matrix element involves a sum over all possible final-states for the residual system.
The transition amplitude in four-point Fermi theory can be written as the contraction of lepton and hadron currents.
The squared amplitude, assuming single boson exchange, can then be written
\begin{align*}
&\delta(E_\nu + M_A - E_\mu - E_N - E_X) \sum_{f} \lvert \mathcal{M}_{fi} \rvert^2 \\
&= G_F^2\cos^2\theta_c \times \\ 
&L_{\alpha\beta}(k_\mu,k_\nu)\left[ \sum_i \rho_i (E_m) H_i^{\alpha\beta}(Q, k_N)\right]_{E_m = E_\nu - E_\mu - T_N - T_X},
\end{align*}
where we factored out the quark-level couplings.
The lepton tensor for charged-current interactions with massless neutrinos is given by
\begin{equation}
L^{\alpha\beta}(k_\nu,k_\mu) = k_{\nu}^\alpha k_\mu^\beta + k_{\nu}^\beta k_\mu^{\alpha} - g^{\alpha\beta} k_\nu\cdot k_\mu - ih\epsilon^{\alpha\beta\gamma\delta} k_{\nu,\gamma} k_{\mu,\delta},
\end{equation}
where $g^{\alpha\beta}$ is the metric tensor, $\epsilon^{\alpha\beta\gamma\delta}$ is the anti-symmetric Levi-Civita tensor, and $h$ is the helicity of the neutrino.

The sum over $i$ includes in principle all interaction mechanisms that may contribute. The function $\rho_i(E_m)$ is the density of states for the mass of the residual system, which depends on the interaction channel.
A microscopic description of the full complexity of all final states that may contribute to the MicroBooNE signal is presently unavailable.
Within the impulse approximation scheme, we include the knockout of nucleons with low momentum and removal energy, which are described within a mean-field picture, as well as a high-momentum background contribution from nuclear short-range correlations. 
We calculate these contributions in the relativistic distorted-wave impulse approximation (RDWIA) to take into account nucleon final-state interactions, in addition to the relativistic plane-wave impulse approximation (RPWIA) and the factorized PWIA, where FSI are neglected.
The description of the hadron tensor in these different approaches is detailed in appendix~\ref{app:hadronT}.
The functions $\rho_i(E_m)$ are determined by considering realistic nuclear spectral functions, and are discussed in the following section.

To include rescattering, events distributed according to the flux-averaged differential cross section of Eq.~(\ref{eq:sigma_fluxfolded}) are generated, and are used as input to several intranuclear cascade models (INCs).
As such we describe the `quasielastic' contribution to the semi-inclusive cross section.

\subsection{The (semi-)exclusive one-nucleon knockout cross section}
\label{sec:Exclusive}
Under exclusive conditions, the cross section can be described with the distorted-wave impulse approximation (DWIA), including an optical potential to take into account FSI.
In this case the imaginary part of the optical potential `absorbs' the flux corresponding to inelastic channels, i.e. it eliminates from the signal those processes in which the outgoing nucleon transfers energy, other than kinetic energy, to the residual system.
Such calculations are commonly used for the description of $(e,e^\prime p)$ experiments~\cite{BOFFI1993}, including the recent analyses of $(e,e^\prime p)$ reactions on argon and titanium~\cite{PhysRevD.107.012005, PhysRevD.105.112002}.
We will use the relativistic DWIA (RDWIA) as a benchmark for the results of the INCs~\cite{Nikolakopoulos:2022qkq}.
The computation of the hadron tensor in the RDWIA is presented in appendix~\ref{app:hadronT}, in the following we discuss the choice of potential.

\subsubsection{The optical potential}
\begin{figure}
\includegraphics[width=0.48\textwidth]{"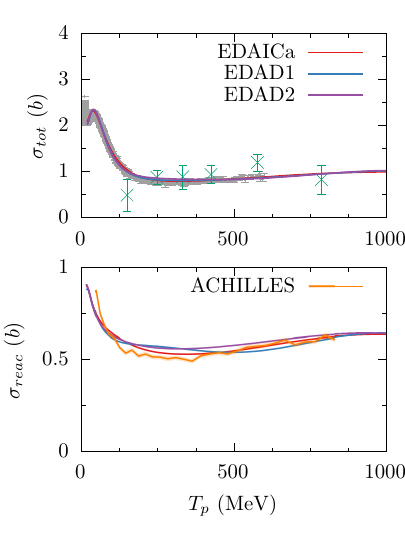"} 
\caption{Total cross section obtained with different optical potentials, compared to neutron calcium (dark-grey) and neutron-argon (green) scattering data (top panel). Reaction cross section for proton-argon scattering (bottom panel). Results obtained with the energy-dependent $A$-independent potential for ${}^{40}$Ca (EDAICa) are shown in addition to those obtained with $A$-dependent (EDAD) potentials from Ref.~\cite{Cooper:1993}. The reaction cross section obtained with \textsc{Achilles} is also shown.}
\label{fig:EDAD_CA}
\end{figure}
We use the energy-dependent $A$-independent optical potential for ${}^{40}$Ca (EDAICa) of Ref.~\cite{Cooper:1993}.
The potential is isospin-independent and  fit to proton-$^{40}$Ca elastic scattering data. The $Z=20$ Coulomb potential was added analytically in the fit of~\cite{Cooper:1993} and it is not part of the extracted potential, therefore, we add a $Z= 18$ Coulomb potential for calculations with ${}^{40}$Ar.
Ref.~\cite{Cooper:1993}, provides alternative energy-dependent $A$-dependent potentials (EDAD), fit to a larger range of targets, and in Ref.~\cite{Cooper:2009} an $A$-dependent `democratic' potential was presented. These were compared to each other and to elastic scattering data off ${}^{40}$Ar in Ref.~\cite{JeffersonLabHallA:2020rcp}, finding that the results with different potentials for $A=40$ practically overlap.
We will use the term ROP (relativistic optical potential) to refer to these potentials.

We show the results for the total neutron scattering cross section obtained with the calcium and A-dependent potentials in Fig.~\ref{fig:EDAD_CA}. 
The top panel shows the comparison to neutron-calcium data, and the recent data for the total neutron-argon scattering cross section measured in CAPTAIN~\cite{PhysRevLett.123.042502}.
The bottom panel of Fig~\ref{fig:EDAD_CA} shows the reaction cross section obtained with different potentials. 
As expected we find that the different potentials give very similar results. All of these potentials are constrained mostly by calcium data for $A=40$. To address possible non-trivial differences between an argon and calcium potential one would need additional argon scattering data or microscopic calculations of argon potentials.

Results for the argon reaction cross section obtained with several intranuclear cascade models were shown in Ref.~\cite{Dytman:2021ohr}. In the bottom panel of Fig.~\ref{fig:EDAD_CA} we include additionally the results obtained with the \textsc{Achilles} cascade~\cite{Isaacson:2022cwh}, which agrees well with the optical model results. Details of the calculation of these latter results are described in Section~\ref{sec:results}.

\subsubsection{Nuclear spectral functions}
The relativistic mean-field (RMF) model of Ref.~\cite{NLSH}, provides the bound state wavefunctions used in the RDWIA calculations.
These are energy-eigenstates of a central potential, which can be labelled by their angular momentum which we denote $\kappa$. Hence for every state $\rho_\kappa(E_m) = \delta(E_m - E_\kappa)$, with $E_\kappa$ the binding energy of the shell.
While this approach can describe discrete excitations of the residual system found at low missing energy~\cite{Lapikas93,Bobeldijk94}, 
at larger missing energy and momentum, the shell model states are found to broaden, with widths of several (up to tens of) MeV~\cite{Johnson88, Tornow90, Atkinson19, BENHAR:RevModPhyseep}. 
Additionally, the shell-model states are found to have reduced occupation numbers, usually of the order of $20-40~\%$~\cite{Udias93,Yasuda10,Giusti11,Atkinson:2018nvp}, which is attributed to long- and short-range correlations (SRC).
The missing strength due to SRC appears at large missing energy and momentum and amounts to approximately  20\%~\cite{SRC_CLAS06,SRC_CLAS18}.

We include the broadening of shell-model states and their reduced occupation in the same way as in Refs.~\cite{Gonzalez-Jimenez:2021ohu, Franco-Patino:2022tvv, Franco-Patino:2023msk}, i.e.,
each RMF state is given a reduced occupancy $N_\kappa$, and the energy is smeared by a Gaussian with width $\sigma_\kappa$. 
The energy density associated with each state is then
\begin{equation}
    \rho_{\kappa}(E_m) = \frac{N_\kappa}{\sqrt{2\pi}\sigma_\kappa } \exp \left\{ - \left(\frac{E_m - E_0}{\sqrt{2}\sigma_\kappa}\right)^2 \right\}.
\end{equation}
 The nucleons missing from the mean-field states are included in a broad $s$-state, to account in an effective way for the high-momentum nucleons in SRC pairs (details in \cite{Gonzalez-Jimenez:2021ohu, Franco-Patino:2022tvv, Franco-Patino:2023msk}). The shape of the resulting momentum distribution and the missing-energy profile are adjusted to reproduce the high $E_m$ and $p_m$ tail of the $^{12}$C Rome spectral function~\cite{Benhar93,Benhar05}, which is based on theoretical predictions and experimental data. The nature of SRC suggests that this shape does not depend on the nucleus~\cite{SRC_CLAS18}, and motivates us to use it for argon. Table~\ref{tab:SFparams} lists the widths, occupation, and missing energy values used for the neutron shells in argon.

To assess the influence of these choices, it is instructive to consider the plane-wave impulse approximation (PWIA), in which it is straightforward to include these effects, by including a realistic hole-spectral function~\cite{BENHAR:RevModPhyseep}.
Indeed, in the PWIA the cross section factorizes as 
\begin{align}
\label{eq:factoriz_SF}
&\frac{\diff\sigma(E_\nu)}{\diff p_\mu \diff \Omega_\mu \diff \Omega_N \diff p_N} = \nonumber \\  &\frac{G_F^2\cos\theta^2_c}{ (2\pi)^2} \frac{p_\mu^2 p_N^2 }{E_\mu E_\nu } L_{\mu\nu}~\frac{M_N^2}{E_N \overline{E}_m} h_{s.n.}^{\mu\nu}(Q, k_N) S(E_m, p_m),
\end{align}
where $S(E_m,p_m)$ is the hole spectral function, and $h_{sn}^{\mu\nu}(Q,k_N)$ is a single-nucleon hadron-tensor. Here $\overline{E}_m = \sqrt{p_m^2 + M_N^2}$. 
Derivation of this expression, the single-nucleon hadron tensor and further discussion can be found in appendix~\ref{app:hadronT}.
Following the arguments of appendix~\ref{app:hadronT}, if we use the PWIA with the RMF bound states described above, the mean-field contribution to the hole spectral function is
\begin{equation}
    \label{eq:SF_factorized}
    S(E_m, p_m) = \sum_\kappa \rho_\kappa (E_m) n_\kappa( p_m).
\end{equation}
Where $n_\kappa$ is the momentum distribution of the shell, given by
\begin{equation}
    n_\kappa(\lvert \mathbf{p}\rvert) = \frac{ f_\kappa^2\left(\lvert \mathbf{p} \rvert \right) + g^2_\kappa\left(\lvert \mathbf{p} \rvert \right)}{4\pi} 
\end{equation}
with $g_\kappa$ and $f_\kappa$ upper- and lower-component radial wavefunctions in momentum space. 

The dependence on energy and momentum in this spectral function is hence factorized. This is a reasonable approach for the mean-field 
contribution, but is in principle not well-suited for the correlated background contribution. Indeed, the latter can be described by considering two-nucleon knockout, and hence the missing energy and momentum dependence are more strongly correlated~\cite{RYCKEBUSCH19961, PhysRevC.89.024603, CiofidegliAtti:1995qe}.

As we will show, because of the flux-averaging, observables are rather insensitive to the energy-dependence of the cross section. Lastly, we stress that this is the representation of the spectral function used in this work in the PWIA, but the cross sections in the RDWIA and RPWIA do not factorize as in Eq.~(\ref{eq:factoriz_SF}).

\begin{table}[b]
\caption{\label{tab:SFparams}%
Parameters for the missing energy profile used in this work. The first column lists the separation energies $E_\kappa$, which are from Ref.~\cite{JeffersonLabHallA:2020rcp}.
Values for the outer shells are compatible with the results from Ref.~\cite{PhysRevD.107.012005}, only the $s$-shell is significantly more deeply bound. The Gaussian widths $\sigma_\kappa$ are also taken from Ref.~\cite{JeffersonLabHallA:2020rcp}. Occupation numbers of the shells $N_\kappa$ are from~\cite{PhysRevC.85.065501}, and are shown as a fraction of completely filled orbitals.
}
\begin{ruledtabular}
\begin{tabular}{lccc}
\textrm{Shell ($nl_J$)}&
\textrm{$E_{\kappa}$ (MeV)}&
\textrm{$\sigma_\kappa$ (MeV)}&
\textrm{$N_\kappa / (2J+1)$}\\
\colrule
$1s_{1/2}$ &  62 & 15 & 1\\
$1p_{3/2}$ &  40   & 8 & 0.95\\
$1p_{1/2}$ &  35   & 8 & 0.95\\
$1d_{5/2}$ &  15.46 & 4 & 0.8\\
$1d_{3/2}$ &  12.84 & 2 & 0.82\\
$2s_{1/2}$ &  12.21 & 2 & 0.85\\
$1f_{7/2}$ &  11.45 & 2 & 0.205\\
\end{tabular}
\end{ruledtabular}
\end{table}

\begin{figure}
\includegraphics[width=0.4\textwidth]{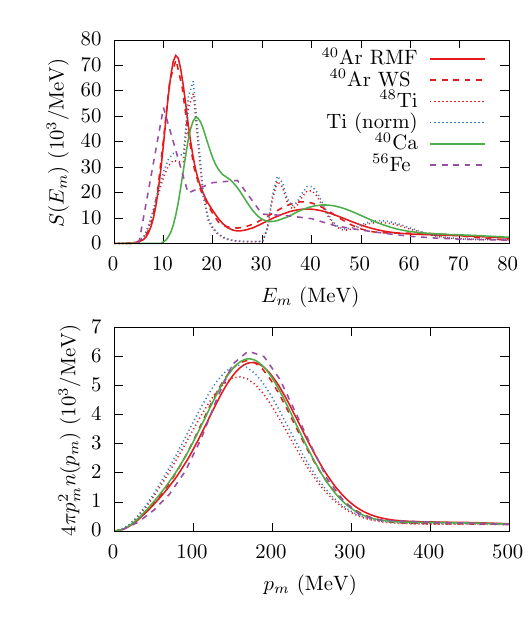}
\caption{Missing energy (top panel) and missing momentum (bottom) distributions obtained with different spectral functions. Results normalized per nucleon.}
\label{fig:dEm_dPM}
\end{figure}

In order to estimate the sensitivity of observables to the choice of spectral function, we compare results obtained with several spectral functions within the PWIA.
This includes a spectral function constructed as in Eq.~(\ref{eq:SF_factorized}), but using instead Woods-Saxon single-particle orbitals to describe the different shells, which have been built using the formalism discussed in Ref.~\cite{Schwierz:2007ve}. This spectral function uses the same widths and occupation numbers as the `test spectral function' used in the $(e,e^\prime p)$ analysis of Ref.~\cite{PhysRevD.105.112002}.
In this case, the correlated part of the spectral function is not of the factorized form of Eq.~(\ref{eq:SF_factorized}).
The background contribution has been obtained following the Ciofi degli Atti parametrization discussed in Ref.~\cite{CiofidegliAtti:1995qe}.
 We express the correlation part as a convolution integral involving the momentum distributions of the relative and center-of-mass motion of a correlated proton-neutron (pn) pair, the different coefficients used are the ones reported in Ref.~\cite{PhysRevD.105.112002}.

Additionally we have considered a reproduction of the proton spectral function fitted to $(e,e'p)$ data for ${}^{48}$Ti of Ref.~\cite{PhysRevD.107.012005}, again with Woods-Saxon single-particle orbitals for the shells. 
In this case, an asymmetric Maxwell-Boltzmann distribution function has been used for the energy density of the shells,
\begin{equation}
    \rho_\kappa(E_m) = \frac{4 x^2}{\sqrt{\pi} \sigma_\kappa }~\exp \left\{ - x^2 \right\}~\theta\left( x\right),
\end{equation}
with $x = (E_m - E_\kappa)/\sigma_\kappa + 1$, and $\theta$ the step function.
In addition to these spectral functions for argon, we consider the calcium and iron spectral functions of Ref.~\cite{Benhar93}. 
The latter are included to obtain energy-dependencies which present a large deviation from the argon and titanium results.

The missing energy and momentum distributions, normalized per target nucleon $N$,
\begin{equation}
S(E_m) = \frac{4\pi}{N} \int  p_m^2\mathrm{d} p_m S(E_m, p_m), 
\end{equation}
and,
\begin{equation}
n(p_m) = \frac{1}{N}\int \mathrm{d} E_m S(E_m, p_m),
\end{equation}
 obtained from these different spectral functions are shown in Fig.~\ref{fig:dEm_dPM}.
The integral of the titanium spectral function yields approximately $20.3$ neutrons, instead of $22$, as can be seen by the sum of occupation numbers in Ref.~\cite{PhysRevD.107.012005}.
We therefore also show the distributions normalized to this reduced number of nucleons in Fig~\ref{fig:dEm_dPM}. This spectral function yields a momentum distribution that is shifted with respect to the others.

It is clear that these different spectral functions represent large variations for the missing-energy dependence. We will show that observables computed in this work are largely insensitive to these variations.
On the other hand, some observables are sensitive to the missing momentum distribution. It is seen however that, with exception of the titanium spectral function, the missing momentum distribution is fairly universal. We use here central values for the titanium spectral function, taking into account the uncertainty shown in Ref.~\cite{PhysRevD.107.012005} would make it compatible with the other results however.

\subsection{Intranuclear cascade models} 
\label{sec:INCs}
Calculations of the exclusive cross section for 1-nucleon knockout, described in the previous section, do not require an explicit description of the residual hadron system. 
Indeed, in the exclusive case, inelastic FSI processes do not contribute to the signal, and the strength lost to such channels is absorbed by the optical potential.
The description of the hadron system in neutrino experiments is more challenging. This lost strength, that leads to more complex final-state configurations, has to be accounted for. 
To do this the intranuclear cascade model (INC) is used.

Without considering the inner-workings of any specific INC, one can formulate the INC approach as a classical approximation to the scattering problem.
Consider a specific exclusive final-state $\lvert X \rangle$, the INC approach can then be understood as the following set of approximations
\begin{align}
\label{eq:INCc}
\lvert \mathcal{M}\rvert^2 &\approx \lvert \sum_{\alpha} \langle \Psi_{0} \rvert T_{1b}  \lvert \psi_{\alpha} \rangle \langle \psi_{\alpha} \lvert X \rangle \rvert ^ 2, \nonumber \\
&\approx \sum_{\alpha}\lvert \langle \Psi_{0} \rvert T_{1b}  \lvert \psi_{\alpha} \rangle \rvert^2 \lvert\langle \psi_{\alpha} \lvert X \rangle \rvert ^ 2 \nonumber \\
&\approx \sum_{\alpha}\lvert \langle \Psi_{0} \rvert T_{1b}  \lvert \psi_{\alpha} \rangle \rvert^2 P(X \lvert \alpha) .
\end{align}
The first line corresponds to introducing an intermediate set of states with quantum numbers $\alpha$, and truncating the operator to a 1-body operator.
The second line is the classical approximation, where the square of the product of amplitudes is replaced by a product of squared amplitudes.
Finally, the last line introduces the conditional probability of producing $X$ from a hadronic final-state specified by $\alpha$.
This last probability is provided by the INC.
It is important to distinguish $P(X \lvert \alpha)$ from $\lvert\langle \psi_{\alpha} \lvert X \rangle \rvert ^ 2$, as the INC does not generally consider the full wavefunction, or full set of quantum numbers $\alpha$.
In the INCs considered in this work, the input is a nucleon with a specific position and momentum. 
The set of quantum numbers for the one-particle-one-hole system is generally more comprehensive. In a spherically symmetric system $\alpha = (\mathbf{p}_N, E_m, \lvert\mathbf{p}_m\rvert, \kappa, m_j, s_N)$, which in addition to the momentum of the nucleon (and it's direction with respect to the lepton system) also includes the missing energy and momentum. The spin of the nucleon ($s_N$) and angular momentum of the residual system ($\kappa, m_j$) are in principle averaged over. 

Apart from this, as interference effects are lost in this approach, a loss of coherence on a length scale of inter-nucleon separation is necessary. 
As such, the approach is expected to break down for low-energy nucleons, whose wavelength can be of the order of the size of the nucleus. Apart from such heuristic arguments, a kinematic region in which this classical approximation is strictly valid is currently not well established.

\subsection{Input to the INC and the inclusive cross section}
\label{sec:inputinclusive}
It is clear to see that the inclusive cross section in the INC approach will be completely determined by the first factor in Eq.~(\ref{eq:INCc}), as $\sum_{\beta} P(\beta | \alpha) = 1$, upon summation over all possible final-states $\beta$.

To obtain an inclusive cross section, one should not include an absorptive optical potential as in the exclusive case.
The most straightforward approach is the (R)PWIA, in which the final-state potential is omitted completely.
This proves to be a reasonable approximation at sufficiently large energy and momentum transfer to the nucleus which for flux-averaged cross sections means away from forward scattering angles~\cite{Gonzalez-Jimenez19}.
It is well known however, that including a modified dispersion relation of the nucleon, i.e. a potential, is crucial to obtain agreement with inclusive $(e,e')$ data~\cite{Ankowski:betterE,Gonzalez-Jimenez19}.
The (R)PWIA cross section can be corrected by including the effect of a final-state potential effectively as in Refs.~\cite{Ankowski:betterE, Benhar91, Benhar08}. This procedure corresponds to a convolution of the double-differential inclusive cross section, i.e. obtained after integration over the nucleon kinematics. 
It is thus not suitable when the full five-fold differential cross section is considered.   
To take into account the final-state potential fully, the (R)DWIA should instead be used. The most natural choice is to compute the final-state wavefunctions as scattering states in the same potential as used for the bound states.
This approach leads to  a correct treatment of Pauli-blocking~\cite{Nikolakopoulos19,Gonzalez-Jimenez19,Nikolakopoulos:2020fti}, satisfies vector current conservation~\cite{Nikolakopoulos:2020fti}, and results in a reasonable description of $(e,e')$ data at low-$q$~\cite{Gonzalez-Jimenez19}.

At higher nucleon energies however, the (energy independent) RMF potential used for the final-state wavefunctions is found to be too strong, and the description of $(e,e')$ data deteriorates~\cite{Gonzalez-Jimenez19}. 
A realistic final-state potential is necessarily energy-dependent (and non-local)~\cite{Dickhoff2017, FESHBACH1958357, Hebborn:2022vzm}.
The most comprehensive method is the relativistic Green's function approach~\cite{Capuzzi91,Meucci03,Meucci05,Ivanov16b}, which consistently accounts for the flux going to inelastic channels in an optical potential.
A simpler but equally effective approach is to neglect the imaginary part of the optical potential when computing the inclusive cross section~\cite{Kim03,Kim07,Butkevich07,Maieron03,Meucci09,Caballero05,Gonzalez-Jimenez20}.
The downside of these above approaches is that the consistency between initial and final state wavefunctions is lost. This is relevant at low momentum transfers or, to be more precise, when the momentum of the struck nucleon is small, in which case the overlap between initial and final state does not vanish and leads to spurious contributions to the cross section. 

To improve the low-energy behaviour, the authors of \cite{Gonzalez-Jimenez19} introduced an energy-dependent real potential, which is by construction identical to the RMF potential (the one used for the bound state) at low nucleon energies and follows the energy dependent behaviour of the ROP at high energies. 
In this way consistency is retained at low energies while the potentials are softer for increasing energies.
We refer to it as the energy-dependent relativistic mean-field (EDRMF), which we use in this work.

\section{Results}
\label{sec:results}
In the following, we compare the exclusive calculations to the equivalent contribution obtained with the INC, and discuss the results obtained for MicroBooNE flux-folded cross sections.
Unless stated otherwise, results include the kinematic cuts used in the MicroBooNE analysis~\cite{MicroBooNE:ddifshort2023}.
This means the muon momentum is restricted to $100~\mathrm{MeV} < p_\mu < 1200~\mathrm{MeV}$, and the event requires one and only one proton with momentum $300~\mathrm{MeV} < p_p < 1000~\mathrm{MeV}$. Any number of additional protons is allowed outside of this range.
Events with charged pions with momenta larger than $70~\mathrm{MeV}$ are excluded from the selection, and events with neutral pions of any momentum are also rejected. Any number of neutrons are allowed.

The MicroBooNE collaboration performed the first double differential measurement on argon in terms of variables that measure the transverse kinematic imbalance.
These are the magnitude of the missing momentum transverse to the beam,
\begin{equation}
\delta P_T \equiv \lvert \np_{p,T} + \np_{\mu,T} \rvert = \lvert \np_{m,T} \rvert,
\end{equation} 
and it's angle with respect to the muon direction,
\begin{equation}
\alpha_T \equiv \acos\left( \hat{\np}_{m,T} \cdot \hat{\np}_{\mu,T} \right),
\end{equation}
along with 
\begin{equation}
\phi_T  \equiv \acos \left( \hat{\np}_{\mu,T} \cdot \hat{\np}_{p,T} \right).
\end{equation}
Where the subscript $_{T}$ denotes the component transverse to the beam direction and $\hat{\np} = \np/\lvert \np \rvert$.

In sec.~\ref{sec:SFcomp} we gauge the sensitivity of the observables to variations in nuclear hole-spectral functions using the factorized PWIA. 
We then use events obtained with the RPWIA and the RDWIA as input to the INCs implemented in \textsc{Achilles}, \nuwro, NEUT, and \incl.
In sec.~\ref{sec:INC_ROP}, we compare the subset of events which do not undergo inelastic FSI (i.e. which are unaffected by the INC), with the results obtained with the Relativistic Optical Potential (ROP).
Finally, in sec.~\ref{sec:Micro_comp}, we compare to MicroBooNE data. We show the effect of using RDWIA events as input compared to RPWIA results.
We discuss the contribution of events that undergo inelastic FSI to the observables.

\subsection{Variation in the spectral function}
\label{sec:SFcomp}
To estimate the sensitivity to our particular choice of the spectral function, we have computed the observables in this work in the PWIA with the alternative spectral functions discussed in sec.~\ref{sec:Exclusive}.

\begin{figure}
\includegraphics[width=0.49\textwidth]{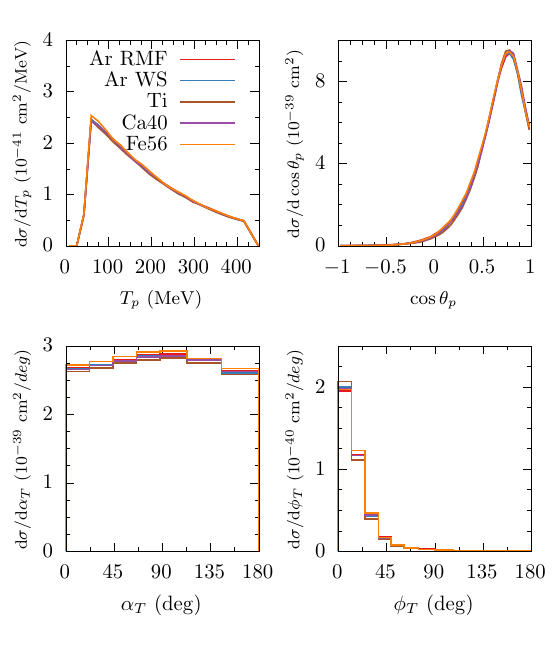}
\caption{Single-differential cross sections obtained with different spectral functions in the PWIA, without FSI. All results are normalized to the number of target neutrons. The results obtained with the titanium spectral function are normalized to $20.3$ neutrons.}
\label{fig:multi_SF}
\end{figure}

Results for single-differential flux-averaged cross sections, normalized per target neutron, are shown in Fig.~\ref{fig:multi_SF}. The titanium results are normalized to the reduced number of neutrons implied by the normalization of the spectral function as discussed before. All these results are obtained without rescattering. 
Similar results are found for the distribution of lepton energies and angles.
We find that most observables are not very sensitive to the rather large variations in the energy distributions shown in Fig.~\ref{fig:dEm_dPM}. 
This is in line with the recent study of Ref.~\cite{Franco-Patino:2023msk}, which considers large variations in the argon spectral function.
It is indeed expected that observables that don't introduce any correlation between lepton and hadron variables, are not very sensitive to the missing momentum or energy distribution, especially in an energy-averaged signal.

\begin{figure}
\includegraphics[width=0.4\textwidth]{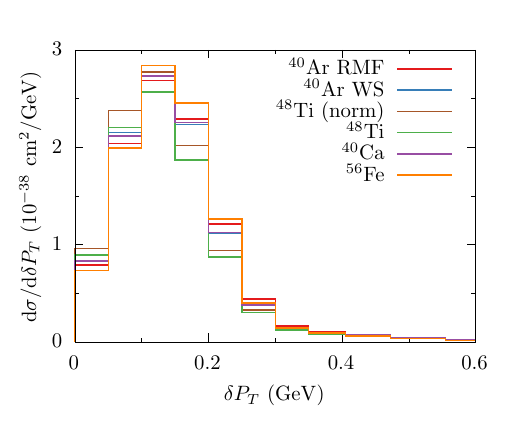}
\caption{$\delta P_T$ distribution obtained with different spectral functions in the PWIA, without FSI. All results are normalized to the number of target nucleons. The line labelled ${}^{48}$Ti (norm) is instead normalized to $20.3$ nucleons.}
\label{fig:dPT_SF}
\end{figure}

The distribution of $\delta P_T$, shown in Fig.~\ref{fig:dPT_SF}, is sensitive to the differences in the missing momentum profiles.
The results obtained with the Woods-Saxon and RMF spectral functions are both very similar to the calcium results.
The clear outlier are the results computed with the titanium spectral function. We also show results normalized to 22 neutrons, instead of the approximately 20.3 obtained by integrating the spectral function. In this case, we find a shift in the $\delta P_{T}$ distribution.
We do not take into account the uncertainty reported on the parameters of the titanium spectral function, given the results shown in Ref.~\cite{PhysRevD.107.012005}, one should assume that when such an uncertainty is taken into account the results are compatible.

Based on these results we can conclude, as expected, that these flux-averaged cross sections are not sensitive to large variations in the missing energy profiles.
However, for neutrino energy reconstruction the distribution of missing energy is of course important, see in particular Refs.~\cite{Gonzalez-Jimenez:2021ohu, VOrden2019}.

The results for $\delta P_T$ are sensitive to the missing momentum distribution. However the latter are almost universal in the spectral functions considered here, as discussed in sec.~\ref{sec:Exclusive}. 
As such the variations in the $\delta P_T$ results, are expected to be small in comparison to experimental uncertainties.
Of course,  realistic momentum distributions can be distinguished from RFG-like or other approximate models through the $\delta P_T$ distribution~\cite{Bathe-Peters:2022kkj, Lu:2015tcr, Dolan:2018zye, MicroBooNE:2023krv}.

\subsection{Comparison of intranuclear cascade models and the optical potential}
\label{sec:INC_ROP}
We generate events according to Eq.~(\ref{eq:sigma_fluxfolded}) and use the outgoing proton momentum as the input to several INCs.
As in Ref.~\cite{Nikolakopoulos:2022qkq}, we benchmark the results of the INCs with a calculation using the relativistic optical potential (ROP).
This comparison is valid for the events that do not undergo inelastic FSI in the INC, i.e. that do not exchange energy with the residual system. 
We select this subset of events to compare to the ROP calculations.

We consider as input to the INCs events distributed according to the 
RDWIA cross section with the real EDRMF potential.
We use the energy-dependence of the RMF-based spectral function. 
Based on the results of the previous section, the results should be largely insensitive to this particular choice of spectral function.
The difference between using the RPWIA or PWIA as input (see appendix~\ref{app:hadronT}), is found to be negligible for these comparisons.

\subsubsection{NEUT}
\begin{figure*}
\includegraphics[width=0.98\textwidth]{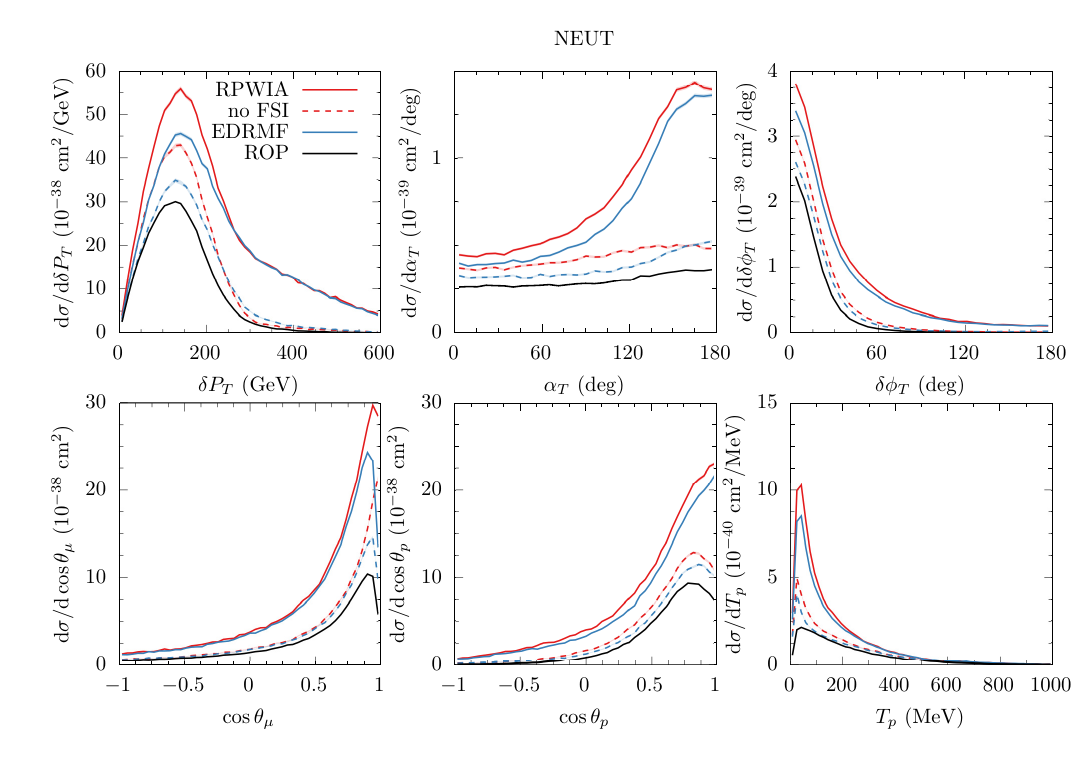}
\caption{Comparison of cross sections without experimental cuts on proton kinematics. The solid lines are the full results including the INC. The results using the RPWIA and EDRMF as input are given in red and blue respectively. The dashed lines of the same colors correspond to the subset of events that are unaffected by the INC, these are labelled 'no FSI', meaning no inelastic FSI in the INC. The solid black lines are the ROP results which can be compared to the 'no FSI' results.}
\label{fig:NEUT_compROP}
\end{figure*}
A similar study of the NEUT INC was presented for T2K flux-folded observables in Ref.~\cite{Nikolakopoulos:2022qkq}.
The inner workings of the NEUT cascade model are described in Refs.~\cite{Hayato:NEUT, Hayato:2021heg}.
The approach is based on the works by Bertini~\cite{PhysRevC.6.631}, but do not include the decays ('evaporation') of the residual system, nor the propagation within a potential. We insert the nucleons in the cascade according to the nuclear density of the RMF model although any realistic density should be suitable as discussed in Ref.~\cite{Nikolakopoulos:2022qkq}. 

We show the results obtained with the NEUT cascade model for both EDRMF and RPWIA inputs in Fig.~\ref{fig:NEUT_compROP}.
These calculations are flux-averaged and with $0.1~\mathrm{GeV} < p_\mu < 1~\mathrm{GeV}$ as in the MicroBooNE data, but do not include kinematic cuts on the final-state hadron system.
We show the total cross section, which includes rescattered events by solid lines. These consider the most energetic proton in the event to define the kinematics. 
The effect of using plane-waves in lieu of distorted waves is significant in the regions of small muon angles, low $\delta P_T$, and for low nucleon energies.
These differences are retained after propagation through the INC. 
The strength at large $\delta P_T$ and $\phi_T$ are completely generated by rescattering in the INC, and are seen to be the same for the RPWIA or EDRMF inputs.
This is because these contributions correlate strongly with large $T_p$ and $\cos\theta_\mu$, where the RPWIA and EDRMF become approximately equal for flux-averaged results.

The subset of events that are unaffected by the INC are shown by dashed lines.
In the NEUT cascade, the propagating nucleon only interacts with the constituent nucleons of the nucleus, and these interactions necessarily knockout additional nucleons or produce mesons. As such the set of events that do not undergo rescattering (dotted lines) are the same as events with 1 and only 1-proton, and no other particles in the final-state. This subset can be compared to the ROP calculations.

We find similar results for the distribution of $T_N$ as in Ref.~\cite{Nikolakopoulos:2022qkq}. The NEUT cascade becomes comparable to the ROP calculation at large $T_N$ but overpredicts by a factor two at low energies. Again, the results that use the RDWIA as input are closer to the ROP results. This is reflected in the $\cos\theta_\mu$ distribution, where forward angles correlate with low nucleon energies.

\subsubsection{NuWro}

\begin{figure*}
\includegraphics[width=0.98\textwidth]{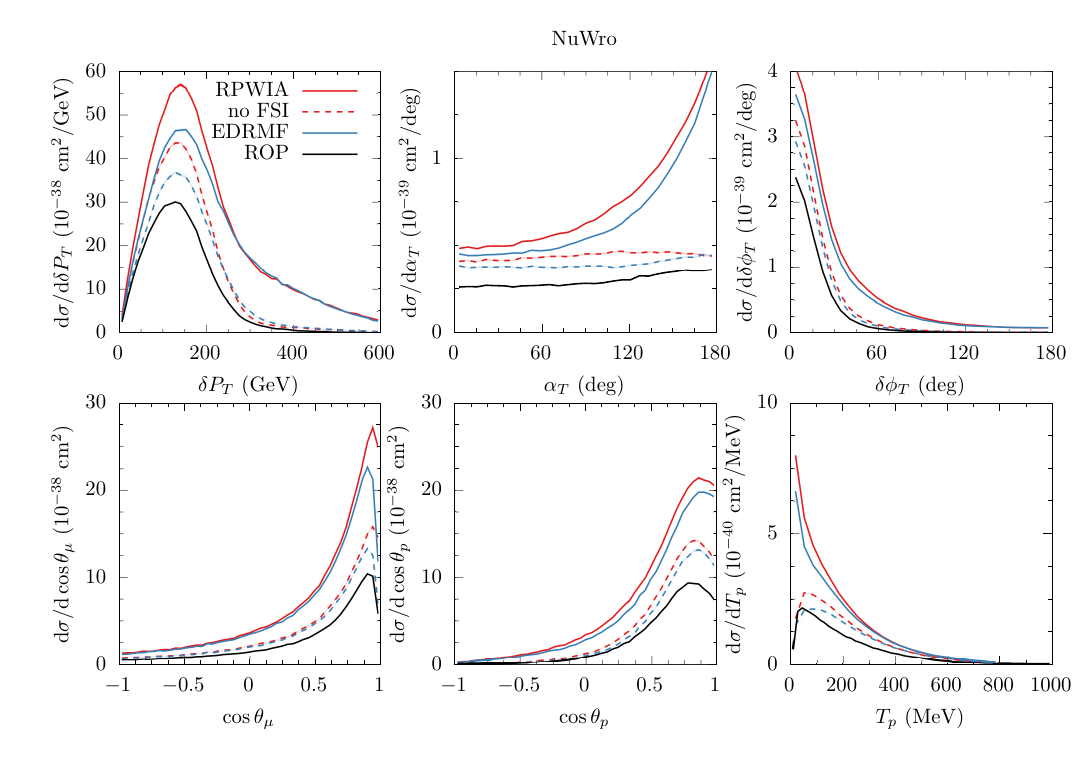}
\caption{Same as in Fig.~\ref{fig:NEUT_compROP}, but for \nuwro.}
\label{fig:NuWro_compROP}
\end{figure*}

The \nuwro INC is primarily based on the seminal papers by N. Metropolis \textit{et al.}~\cite{Metropolis:1958wvo,Metropolis:1958sb} but includes input from up-to-date physical data and treatment of additional effects present in the nuclear medium. The model describes the in-medium propagation of nucleons~\cite{Niewczas:2019fro,Golan:2012wx}, pions~\cite{Golan:2012wx}, and hyperons~\cite{Thorpe:2020tym}. It employs the space-like approach: sampling Monte Carlo steps using the classical formula for passing a distance $\Delta x$ with no re-interactions
\begin{equation}
    P(x) = \exp(-\Delta x / \lambda),
\end{equation}
where $\lambda = (\rho \sigma)^{-1}$ is the mean free path calculated locally, expressed in terms of the nuclear density $\rho$ and the effective interaction cross section $\sigma$. A given particle interacts if its sampled step is not greater than the maximal value of $0.2 \ \mathrm{fm}$.
The energy and momentum of nucleon targets, are determined from the local Fermi gas model.

A given run of the INC concludes when all the moving hadrons either exit the nucleus or lack sufficient kinetic energy to escape the nuclear potential (with the separation energy of $7 \ \mathrm{MeV}$). In \nuwro, the de-excitation process of the remaining nucleus is not modeled.

\nuwro employs several effective methods to account for in-medium corrections to the propagation of hadrons within the INC model. 
Interactions are allowed only if the selected final-state kinematics brings all involved nucleons above the Fermi sea. 
As described in Ref.~\cite{Niewczas:2019fro}, nucleon-nucleon cross sections are quenched according to the procedure of Ref.~\cite{Pandharipande:1992zz} for the elastic and the parametrization of Ref.~\cite{Klakow:1993dj} for the inelastic processes. The pion-nucleon cross sections are evaluated using the model of L. L. Salcedo \textit{et al.}~\cite{Salcedo:1987md}, which includes the effect of the in-medium modification of the $\Delta$-resonance self-energy, see Ref.~\cite{Golan:2012wx}. Finally, we introduce the impact of nucleon-nucleon correlations into the effective density profile using a custom model based on Variational Monte Carlo calculations of two-body nucleon densities by R. B. Wiringa \textit{et al.}~\cite{Pandharipande:1992zz,Wiringa:2013ala}. Such holes in nuclear density around the starting point of nucleon propagation result in a reduced probability of nucleon-nucleon interactions over short distances, leading to an increase in the predicted values for the $(e,e^\prime p)$ nuclear transparency~\cite{Niewczas:2019fro}. One can also obtain a similar effect while using the formation zone/time for nucleons entering the INC~\cite{Golan:2012wx}, which, in the recent \nuwro versions, is retained only for the inelastic neutrino-nucleus interactions.

We present the results obtained with \nuwro version 21.09 in Fig.~\ref{fig:NuWro_compROP}. 
In \nuwro, 'absorption' can occur for low-energy nucleons. This takes into account collective excitations effectively. The absorbed energy would lead to an excited compound system, which has to be decayed accordingly as studied in Ref.~\cite{Ershova:2022jah}.
Here, we do not include such deexcitations, and as such, this absorption can produce 0-nucleon final states and final states with 1 proton, which underwent rescattering. This is in contrast with the NEUT results, where all 1-proton events do not suffer rescattering.
We find that \nuwro provides a smaller cross section than NEUT at low $T_N$, where it is very much in line with the ROP results. 
This is due to the low probability for low-energy protons to be affected by FSI in NEUT, which we further discuss in Sec.~\ref{sec:incs_comparison}.
In \nuwro a larger fraction of high-$T_N$ nucleons are unaffected by FSI, because of the effect of short-range correlations on the mean-free path as discussed above and in Refs.~\cite{Niewczas:2019fro, PhysRevC.45.791}. 
This explains why the contribution of events that do not undergo FSI is larger in \nuwro than in the other INCs we consider.
We use this version of \nuwro throughout this work, as it is the standard implementation.
In appendix~\ref{app:alternate}, we show the results obtained with \nuwro when the effect of SRC on the mean free path is not included.
In this case, for most observables, the EDRMF results become comparable to the ROP.

\subsubsection{\textsc{Achilles}}
\begin{figure*}
\includegraphics[width=0.98\textwidth]{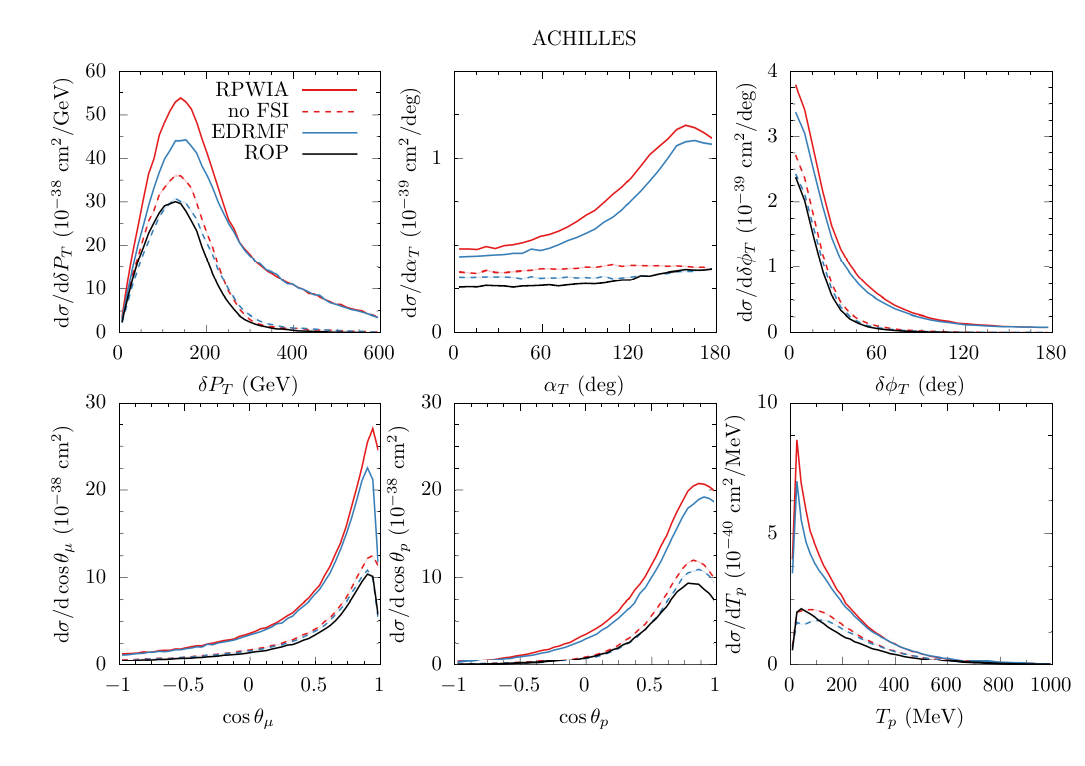} 
\caption{Same as in Fig.~\ref{fig:NEUT_compROP}, but for \textsc{Achilles}.}
\label{fig:ACHILLES_compROP}
\end{figure*}
The INC used in \textsc{Achilles}~\cite{Isaacson:2022cwh} is described in Ref.~\cite{Isaacson:2020wlx}.
The approach is different from the NEUT and \nuwro INCs in which the mean-free path is determined by the in-medium cross section and the nuclear density.
\textsc{Achilles} uses configurations of nucleons (i.e. statistical samples of nucleon positions) as seeds for the cascade.
The interaction probabilities are functions of the impact parameter (distance of closest approach between the propagating particle and a background nucleon) and the total cross section. There are currently two different models for this parameter called Cylinder and Gaussian (see Ref.~\cite{Isaacson:2020wlx} for more details).
Nucleon-nucleon interactions can occur over a range which becomes larger when the in-medium cross section grows. In this work we use the Cylinder interaction model, the NASA parametrization~\cite{NASAinteractions} for the nucleon-nucleon cross section, and the in-medium correction to the nucleon-nucleon cross section of Ref.~\cite{PhysRevC.45.791}.

We show in Fig.~\ref{fig:ACHILLES_compROP} the comparison of these event selections in \textsc{Achilles} with the results obtained with the relativistic optical potential. 
The results obtained with the EDRMF inputs are found to be comparable to the ROP results for many observables, $\delta P_T$, $\delta \phi_T$, and $\cos\theta_\mu$ in particular.
The main differences are exposed as function of $T_p$, where it is seen that the large $T_p$ cross section in \textsc{Achilles} is larger, while at low-$T_p$ the cross section is slightly smaller. 
The correlation between high-energy nucleons and forward scattering angles then explains also the result for $\cos\theta_p$. 

In this work, we use configurations for the INC obtained from RMF calculations for argon.
In contrast to the configurations obtained from quantum Monte-Carlo for carbon used in Ref.~\cite{Isaacson:2020wlx}, the present configurations do not feature a reduction of the two-nucleon density at short distances due to correlations.
As discussed in Ref.~\cite{Isaacson:2020wlx}, this effect is expected to be small for these observables. This is partly because the \textsc{Achilles} INC allows for interactions over a distance, thus reducing sensitivity to the local density. The main reason is because a formation-time, during which the nucleon is not allowed to reinteract, is included after any interaction~\cite{Landau:1953gr, Golan12}.
This formation time hence means the mean-free path is increased, and is not sensitive to the short-distance structure.
We have checked this explicitly, by simply considering a subset of RMF configurations where the inter-nucleon distance is larger than 1~fm. 
Similarly, we have used configurations from Woods-Saxon calculations~\cite{Schwierz:2007ve}. The results shown in Fig.~\ref{fig:ACHILLES_compROP} are insensitive to these choices, because of the formation time included.
For completeness, we include in Appendix~\ref{app:alternate} the comparison without including the formation-time effect.

\subsubsection{INCL}
\begin{figure*}
\includegraphics[width=0.98\textwidth]{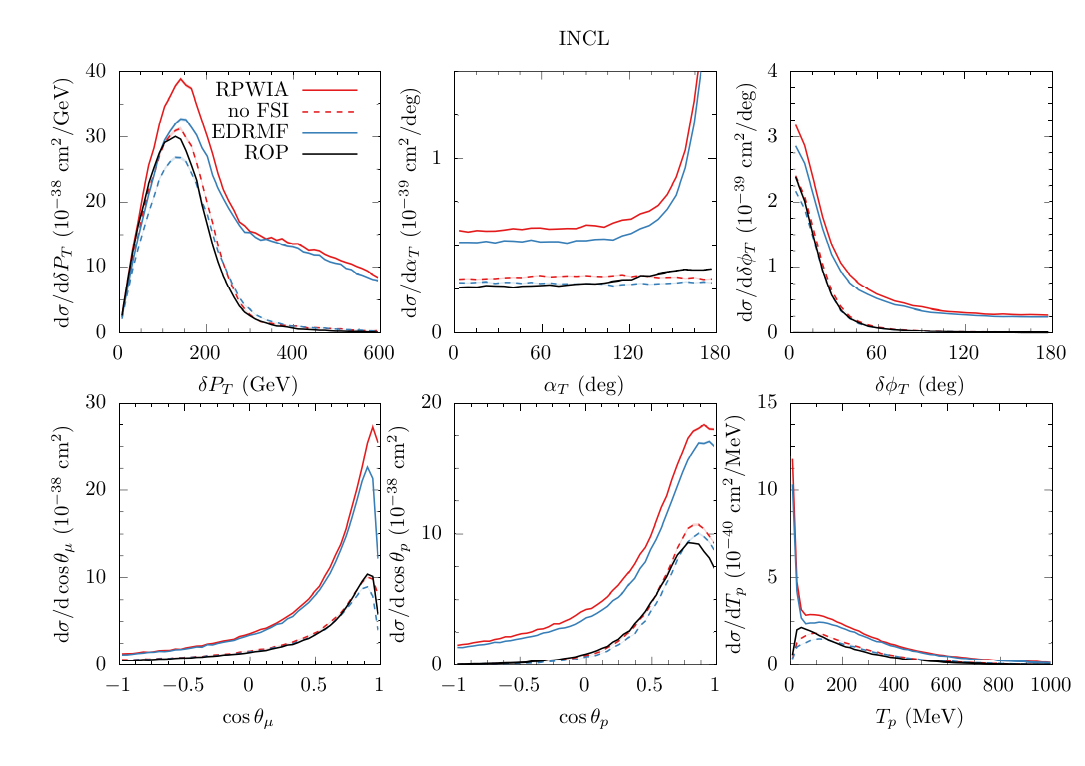} 
\caption{Same as in Fig.~\ref{fig:NEUT_compROP}, but for \incl.}
\label{fig:INCL_compROP}
\end{figure*}
The initial objective of the Liège Intranuclear Cascade Model (\incl) was to simulate nuclear reactions triggered by hadrons and light nuclei in the energy range of tens of MeV to a few GeV. It shows a remarkable agreement with various experimental data, highlighting its reliability and accuracy (as shown, for example, in Refs.~\cite{IAEA} and~\cite{2015}). An essential aspect of \incl is the requirement of a subsequent de-excitation model to properly release the excitation energy accumulated by the nucleus during the cascade process. Among the available models, \abla~\cite{ABLA, Rodriguez-Sanchez:2022Abla} stands out for its proven effectiveness in handling light nuclei, as shown in Ref.~\cite{ABLAlow}. Even though it was not initially designed to simulate neutrino interactions, recent significant applications have shown the importance of nuclear cluster production in neutrino interactions, and the subsequent deexcitation modeling as detailed in Refs.~\cite{Ershova:2022jah, Ershova:2023dbv}.

\incl is a classical model with additional components to simulate quantum effects. This model assigns nucleons specific positions and momenta within Woods-Saxon, modified-harmonic-oscillator (MHO), or Gaussian potential wells depending on the target nucleus~\cite{Rodriguez-Sanchez:2017odk}. In a classical picture, position and momentum have a one-to-one correlation. \incl loosens the correlation and enables nucleons to sometimes extend beyond expected boundaries by employing the Hartree-Fock-Bogoliubov formalism, adding a layer of complexity and realism to the model. Further details on this framework can be found in Ref.~\cite{Rodriguez-Sanchez:2017odk}.

\incl cascade follows the time-like approach, meaning that all the particles propagate until two particles reach the minimal interaction distance, the particle decays, or reaches the border of the potential and attempts to leave. Notably, as particles exit the nucleus, they may cluster with adjacent nucleons, forming nuclear clusters in the process~\cite{Mancusi:2014fba}. For each simulated neutrino event, \incl calculates a chronological table of upcoming events inside the cascade, which includes collision, reflection or transmission at the surface, and decay. Then, as the cascade evolves in time steps given by the duration until the impending interaction, the table is updated with the new possible events. The INC concludes when either there are no more participants left, the mass number of the residual system is less than 4, or the event reaches the stopping time determined by the model.

\incl offers two distinct approaches to Pauli blocking: the strict model, which forbids interaction if the projectile momentum is below the Fermi momentum, and the statistical model~\cite{Henrotte}. The latter includes only nearby nucleons in the phase-space volume and acts according to the calculated occupation probability. In this study, since strict Pauli blocking should be applied to the externally simulated primary vertex, only statistical Pauli blocking will be used for the following interactions. Furthermore, the Coherent Dynamical Pauli Principle (CDPP)~\cite{PhysRevC.66.044615} is employed to avoid problems resulting from the possible creation of holes in the Fermi sea during the initialization of the nucleus.

The results for \incl are shown in Fig.~\ref{fig:INCL_compROP}. 
The set of events not affected by FSI is comparable in shape and magnitude to the ROP results.
In this case, it is seen that \incl features stronger FSI, and the magnitude of the RPWIA input is overall closer to the ROP result.
The overall shape of the total signal is very different in \incl compared to the other INCs.
\incl is seen to produce more events with large $\alpha_T$, $\phi_T$ and $\delta P_T$ than the other INCs.
The $T_p$ distribution of all protons, features a distinctive build-up at low energies, many of these are produced in fragmentation of the residual system.
However, \incl also produces a smaller amount of protons than the other INCs in the region of $50~\mathrm{MeV} < T_p < 200 ~\mathrm{MeV}$, with a larger high-energy tail.

\begin{figure}
\includegraphics[width=0.48\textwidth]{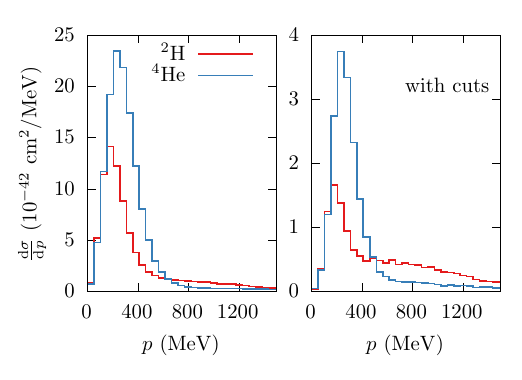} 
\caption{Momenta of helium and deuterium produced in \incl+\abla. The left panel considers the full simulation, while the right panel includes the cuts on protons and pions employed in the MicroBooNE analysis.}
\label{fig:clusters_INCL}
\end{figure}

It is notable that while the particle content produced by the other INCs is only nucleons and mesons, in the \incl the cascade and deexcitation produce a significant number of light nuclear clusters, as well as deexcitation photons. 
The produced photons are all between $4$ and $8$ MeV, purely from deexcitations below the threshold for hadron knockout.
We find that $20\%$ of events contain at least one ${}^{4}$He, $18 \%$ a ${}^{2}$H, and $8\%$ of events contain ${}^{3}$H and ${}^{3}$He, when only the lepton kinematics are restricted to $100 < p_\mu < 1200~\mathrm{MeV}$.
After the cut on proton kinematics and pions employed in MicroBooNE the percentages reduce to $8\%$ of selected events for both helium and deuterium.
The momentum spectrum of produced helium and deuterium is shown in Fig.~\ref{fig:clusters_INCL}.
Deuteron would be easily distinguishable from protons in a LArTPC~\cite{Ershova:2022jah}. An estimate of the detection threshold for deuteron is a momentum of $p > 500~\mathrm{MeV}$ ($p > 1200~\mathrm{MeV}$ for helium)~\cite{Furmanski:private}. 
This threshold is conservative, it is an estimate of the momentum required to leave a $2~\mathrm{cm}$ track, with the stopping power for deuteron in argon taken from the values reported for calcium~\cite{osti_4529550}.
For comparisons, the proton momentum threshold of $250~\mathrm{MeV}$ reported in~\cite{MicroBooNE:2024tmp} corresponds to a track length of around $1~\mathrm{cm}$.
As seen in Fig.~\ref{fig:clusters_INCL}, \incl predicts detectable high-momentum deuterons. As far as we know, explicit measurement of deuterons in LArTPCs have currently not been reported.

\subsubsection{Comparison of INCs}
\label{sec:incs_comparison}
\begin{figure*}
\includegraphics[width=0.32\textwidth]{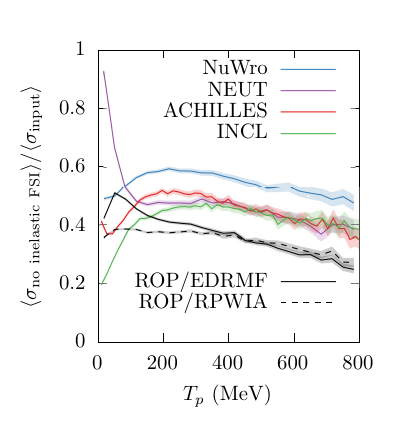}
\includegraphics[width=0.32\textwidth]{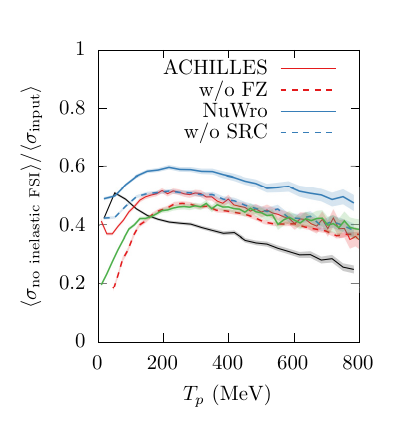}
\includegraphics[width=0.32\textwidth]{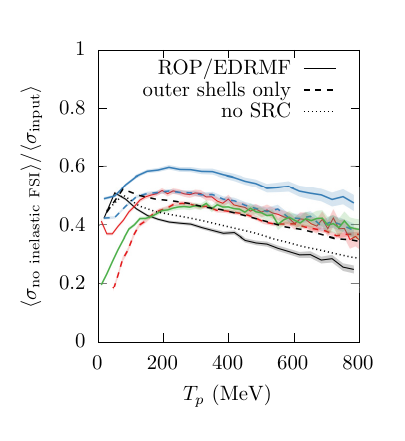}
\caption{Comparison of cross sections as function of $T_p$. We show the ratio of the events that are unaffected by inelastic FSI with respect to the events that are used as input to the INC. In the INC approach these are independent of the input calculation. For the ROP we show the ratio with respect to both the RPWIA and EDRMF calculations. Linestyles repeated in different panels correspond to the same models. The left panel shows the results using the standard parameters for the INCs used throughout this work. The middle panel shows the effect of the formation-zone and treatment of SRCs in \textsc{Achilles} and \nuwro respectively. The rightmost panel shows the ratio ROP/EDRMF when only a limited number of states are taken into account. These corresponds to the four most outerlying shells (dashed black), all shells without the SRC background (dotted black), and the result including the background (solid black).
}
\label{fig:TN_AC_NEUT}
\end{figure*}
In previous sections we discussed in detail the results for the different INCs used in this work for several kinematic variables. 
To compare the different INC results to each other and to the ROP results, the natural variable is the distribution of nucleon energy $T_N$.
Indeed, due to the factorized treatment, Eq.~(\ref{eq:INCc}), where the input to the INC is a nucleon momentum and position, the probability of interaction is completely determined by the magnitude of the momentum after averaging over the position. 
The results for other kinematic variables follow essentially from their correlation with nucleon energy, which are determined by the input calculation.
We show the results for the $T_p$ distribution of events that do not undergo inelastic FSI in the INC, dashed lines in Figs.~(\ref{fig:NEUT_compROP}-\ref{fig:INCL_compROP}), as a ratio to the $T_p$ distribution used as input in Fig.~\ref{fig:TN_AC_NEUT}. 
Due to the factorized treatment, these do not depend on the calculation used as input. Indeed, this is the nuclear transparency obtained in the INC as  in Ref.~\cite{Dytman:2021ohr}, which is not to be confused with the experimental definition of nuclear transparency in $(e,e'p)$ experiments.
The latter is usually obtained as a ratio to a PWIA calculation, and depends on the phase-space under consideration, and the details of the PWIA model, such as the choice of current operator and nucleon form factors.

The different INCs agree with each other for kinetic energies larger than $400~\mathrm{MeV}$, with the exception of \nuwro. This is because of the effect of correlations on the mean-free path as implemented in Ref.~\cite{Niewczas:2019fro}.
If this effect is omitted, as shown by dashed blue lines in the middle panel, the \nuwro results agree with the other INCs.
The INC results can be compared to the ratio of the ROP result with the calculation used as input. Here, the result does depend on the denominator, they are shown as a ratio to both EDRMF and RPWIA.
At high kinetic energies, $T_p > 400~\mathrm{MeV}$, these ratios are the same, indicating similar cross sections in the EDRMF and RPWIA.
The shape obtained in the ROP is the same as the INCs in this regime, but the latter give cross sections that are larger by $25\%$.
The comparison shows that there is no full agreement between any of the INCs and between the INCs and the ROP. 
Good agreement found for other variables shown in Figs.~(\ref{fig:NEUT_compROP}-\ref{fig:INCL_compROP}) arise from the overprediction at large-$T_p$ being cancelled by an underprediction at smaller energies.
The main differences between the INCs are found at low-$T_p$, where none of the models agree.
The NEUT result is clearly the outlier, giving a large rise in transparency at low energies. 

The differences between the other INCs can be partly understood from the treatment of correlations, as we show in the middle panel of Fig.~\ref{fig:TN_AC_NEUT}.
One sees that if the effect of SRC in \nuwro is removed, the \nuwro and \textsc{Achilles} INCs agree to a large degree.
As discussed in Ref.~\cite{Isaacson:2020wlx}, the effect of correlations, which decrease the two-nucleon density at small distances, does not cause a large effect on the transparency in \textsc{Achilles}.
This is because of the inclusion of a formation time, in which a nucleon propagating through the INC does not interact. The formation time removes the sensitivity to the two-nucleon density at short distances, which in \textsc{Achilles} can be explicitly included through the use of configurations of nucleons.
We have confirmed this also for argon, by simply using subsets of the RMF configurations where the nucleon-nucleon separation is large, finding essentially a negligible effect just as in Ref.~\cite{Isaacson:2020wlx}.
On the other hand, we show in the middle panel  of Fig.~\ref{fig:TN_AC_NEUT} the \textsc{Achilles} result in which the formation-zone is not included and using purely mean-field configurations. 
This case is the result for a purely independent particle model, and the transparency becomes similar to the \incl result.

None of these modifications significantly reduce the gap between the ROP and the INC calculations at large energies.
This difference, while smaller than the variations at lower energies, and in a kinematic region that gives a small contribution to the MicroBooNE signal, is interesting because it exposes a systematic disagreement of the ROP and the INCs.
The ROP approach, in this $T_p$ region is used to model the effect of FSI in the $(e,e'p)$ experiments to extract or benchmark the partial occupations in the mean field region.
This disagreement, taken at face value, would imply an inconsistency with the electron scattering results, used to extract the spectral function in the first place.
However, the comparison shown here extends over the whole phase space, probing the nuclear spectral function also at large $E_m$ and $p_m$.
In contrast, $(e,e'p)$ experiments are usually performed at low $E_m$ and $p_m$~\cite{Van-Der-Steenhoven88,Kramer89,Garino92,Leuschner94,Holtrop98,Lapikas00,Dutta03,Fissum04,Rohe05}, including measurements of nuclear transparency (see~\cite{Niewczas:2019fro} for a review). 
Hence, we also perform calculations restricted to the kinematic region $E_m < 30~\mathrm{MeV}$. These include only the four least bound shells listed in Table~\ref{tab:SFparams}, and no high-momentum background.
The ratios of ROP calculation with respect to the EDRMF are shown in the rightmost panel of Fig.~\ref{fig:TN_AC_NEUT}. 
In this case the ratio increases significantly, and the EDRMF result becomes compatible with the high-$T_p$ result obtained in the INCs. 
We find that while the $T_p$ spectrum in EDRMF calculations is similar for different shells, the ROP leads to stronger absorption for the inner, more deeply bound, shells (with larger $E_m$). 
This is seen by considering the result for the ratio where all mean-field states are included, but without the high-momentum background, also shown in the rightmost panel of Fig.~\ref{fig:TN_AC_NEUT}. Inclusion of the deeper shells reduces the ratio. 
This is in accordance with the intuition that nucleons originating from deeper inside the nucleus are more affected by inelastic FSI than those at the surface. 
This is of course also naturally the case in the INCs. 
The present INC results use an averaged density to determine the starting position of the nucleon.
However, nucleons arising from the high-$p_m$ SRC background are absorbed strongly in the ROP calculation due to it being modeled as an  $s$-shell  broad in momentum space, which corresponds to a narrow distribution deep in the nucleus in coordinate space.
To properly include FSI for short-range correlated pairs, both in the INC and ROP, one needs a more detailed model for the phase-space distribution of correlated pairs~\cite{Cosyn:2021ber}.

\subsection{MicroBooNE data}
\label{sec:Micro_comp}
\begin{figure}
\includegraphics[width=0.98\columnwidth]{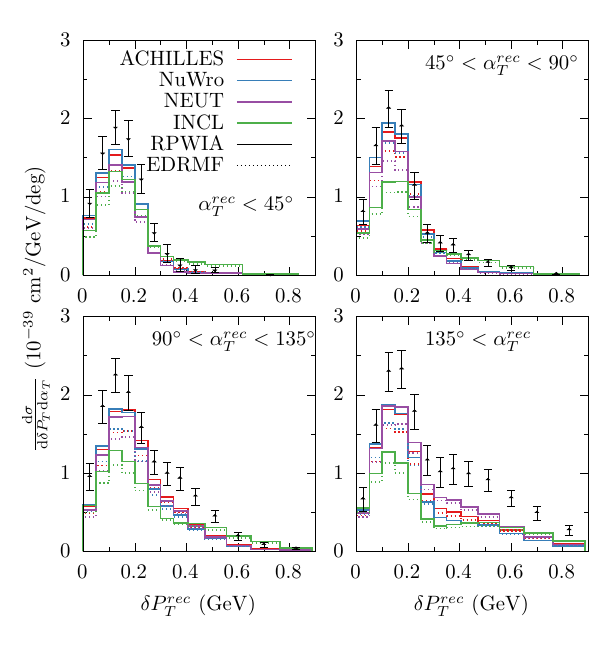}  \\
\caption{Double differential cross sections in terms of reconstructed variables compared to MicroBooNE data. The panels show different bins of $\alpha_T^{rec}$. Different INC models are labelled by different colors. Solid (dashed) lines use the RPWIA (EDRMF) approach as input to the INC.}
\label{fig:double_diff_p_mb}
\end{figure}
We compare the results of different cascade models, using both RPWIA and EDRMF inputs to the experimental data of Ref.~\cite{MicroBooNE:ddifshort2023, MicroBoonE:ddiflong2023}.
It is important to note that the data is presented in terms of reconstructed variables, which we indicate with a superscript ${}^{rec}$.
To compare calculations to the reconstructed data, we perform the multiplication of results binned in true variables with the smearing matrices provided with the data~\cite{MicroBoonE:ddiflong2023}.
This procedure leads to a renormalization and redistribution of events over the different bins, and hence a naive interpretation of the cross section in terms of true kinematic variables is in principle impossible. 

\begin{figure}
\includegraphics[width=0.95\columnwidth]{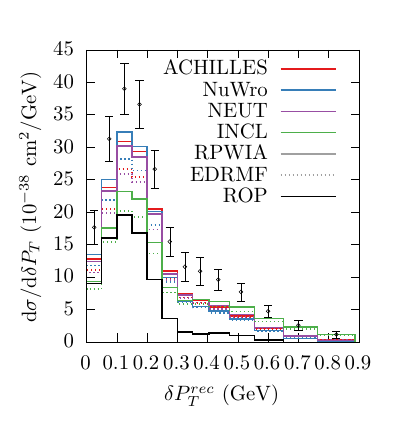} 
\caption{Single differential cross section as function of reconstructed $\delta P_T^{rec}$ compared to MicroBooNE data. The different INCs use a unique color, and solid (dashed) lines show the result using the RPWIA (EDRMF) approach as input. The ROP calculations do not use any INC.}
\label{fig:single_diff_mb}
\end{figure}

We present comparisons to the cross section double-differential in $\delta P^{rec}_T$ and $\alpha^{rec}_T$ in Fig.~\ref{fig:double_diff_p_mb}.
In the absence of FSI, the $\alpha_T$ distribution is quite uniform, as seen in the previous two sections.
Inelastic FSI tend to increase the missing momentum, and redistribute strength from small $\alpha_T$ to large $\alpha_T$, thus small $\alpha_T$ correlates with a smaller contribution of inelastic FSI. This is observed in the comparison, showing a redistribution of strength to high $\delta P_T^{rec}$ with increasing $\alpha^{rec}_T$. 
Similarly, two-nucleon knockout contributions through MECs or SRCs which we do not consider explicitly, contribute mostly at large $\alpha_{T}^{rec}$, hence it is expected that the high-$\delta P_T$ tail is underpredicted.

\begin{figure}
    \centering
    \includegraphics[width=0.95\columnwidth]{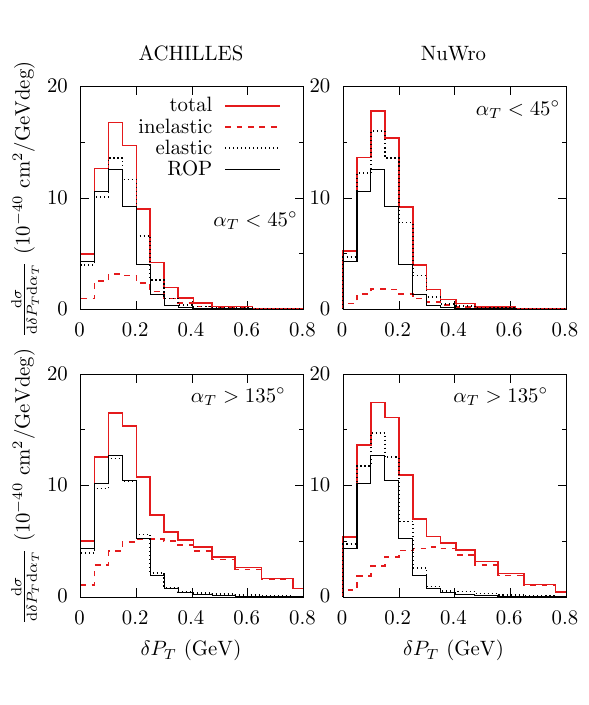}
    \caption{Double differential cross section as function of true kinematic variables, for \textsc{Achilles} (left) and \nuwro (right) INCs. The contribution of events that undergo inelastic FSI are shown in red dashed lines. Although the total strength in the low-$\delta P_T$ region is comparable between the two INCs, the make-up of the signal is not. The dotted black line shows the contribution of events that do not undergo rescattering in the INC, and is compared to the ROP calculation shown by solid black lines.}
    \label{fig:RESC_ACH_NUW}
\end{figure}

The variation of the results with the choice of INC is large, in particular the difference between \incl and the other INCs stands out. 
In the region of small $\alpha_T^{rec}$ and $\delta P_T^{rec}$, the results correlate with the discussion in the previous section, \nuwro providing the least reduction while \incl provides the largest reduction in this mean-field region. 
The \textsc{Achilles}, \nuwro and NEUT results at low-$\delta P_T^{rec}$ are similar in larger $\alpha_T^{rec}$ bins, while \incl predicts a strong reduction of strength compared to the other INCs.
The \incl results have more strength in the high-$\delta P_T^{rec}$ and high-$\alpha_T^{rec}$ regions relative to the mean-field peak compared to the other INCs, with an overprediction of the data at large $\delta P_T$ as a result.

\begin{figure}
\includegraphics[width=0.98\columnwidth]{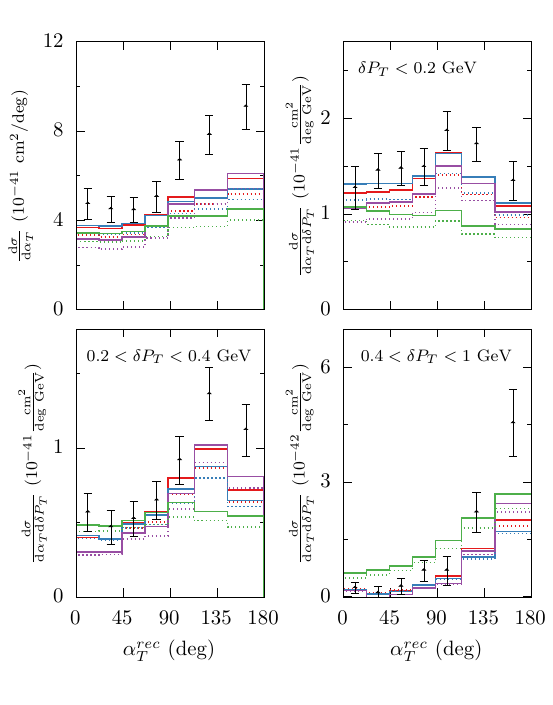}  \\
\caption{Cross sections for MicroBooNE kinematics as function of $\alpha_T^{rec}$. The top left panel is the single-differential cross section, the other panels are double differential in different $\delta P_T^{rec}$ bins. Legend is the same as in Fig~\ref{fig:double_diff_p_mb}.}
\label{fig:double_diff_a_mb}
\end{figure}

The single differential cross section in terms of $\delta P_T^{rec}$ is shown in Fig.~\ref{fig:single_diff_mb}. 
A underprediction of the data is expected, which is most significant for the EDRMF results. Again, \textsc{Achilles}, \nuwro and NEUT give similar results, while \incl gives a significantly smaller cross section at low $\delta P_T^{rec}$.
We show also the result obtained with the ROP, which of course hasn't been passed through the INC, as it includes the necessary absorption.
The ROP is essentially a calculation of the cross section for protons that do not undergo inelastic FSI, and hence provides a theoretical minimum for the nucleon knockout contribution.
The main difference between the ROP and the INC results here comes from the contribution of rescattered events, which also contribute at low $\delta P_T$.
The contribution of rescattering also explains why the \textsc{Achilles} and \nuwro cascades provide similar results, even though the transparency in the latter is larger as shown in Fig.~\ref{fig:TN_AC_NEUT}.
The contribution of rescattering is shown explicitly in Fig.~\ref{fig:RESC_ACH_NUW}.
We show results for \nuwro and \textsc{Achilles} in the most forward and backward $\alpha_T$ bins in Fig.~\ref{fig:RESC_ACH_NUW}. We use true instead of reconstructed variables.
The contribution of events that do not undergo inelastic FSI are shown separately, and can be compared to the ROP calculations.
The underlying differences in the INC predictions are larger than those observed in Fig.~\ref{fig:double_diff_p_mb}. The relative contribution from events that do (not) undergo inelastic FSI is different for both approaches. 
It is found that no combined kinematic cut on $\alpha_T$ and $\delta P_T$ can remove the contribution of rescattering. This contrasts with the results found for the T2K experiment, where the INC and ROP results are similar. This is because for the latter the main effect of FSI is a removal of events from the phase space, protons lose energy and end up below the detection threshold. For MicroBooNE, rescattered protons are non-negligible.

The projection in terms of $\alpha_T^{rec}$ for different slices of $\delta P_T^{rec}$ is shown in Fig.~\ref{fig:double_diff_a_mb}.
The single-differential cross section, shown in the top-left panel, provides the expected picture, namely that the cross section at large $\alpha_T^{rec}$ is underpredicted due to the lack of additional interaction mechanisms.
Interestingly, the deficit seen in the projection onto $\delta P_T^{rec}$, Fig.~\ref{fig:double_diff_p_mb} at low-$\delta P_T^{rec}$ and $\alpha_T^{rec}$ is smaller in this comparison, shown in the top-right panel of Fig.~\ref{fig:double_diff_a_mb}. The deficit remains at low-$\alpha_T^{rec}$ for the single differential cross section.
This shows that one should be careful when interpreting the reconstructed variables.
Again, \incl is the outlier, with a significant shape difference compared to the other INCs.

The results shown in Figs.~\ref{fig:double_diff_a_mb} and~\ref{fig:double_diff_p_mb} indicate that the effect of using the RPWIA as input instead of the full EDRMF calculation is significant. In all double differential bins, and for the single differential cross section shown in Fig.~\ref{fig:single_diff_mb} we find the RPWIA cross section to be $10\%$ larger than the EDRMF.
This reduction is of course consistent between event generators, and tends to be largely independent of the value of $\delta P_T$ and $\alpha_T$.

While the underestimation of the cross section at large $\delta P_T$ and $\alpha_T$ is expected, we find a clear deficit also at low-$\delta P_T^{rec}$ and $\alpha_T^{rec}$, in particular when the more realistic EDRMF is used as input.
One expects meson exchange currents and resonance production to contribute mainly at large $\alpha_T$ and $\delta P_T$, so one might need an increase of the quasielastic contribution in the mean-field region as well.
Such a direct interpretation is not strictly valid as we showed above: there is a model-dependent rescattering contribution, and the reconstructed variables in the projections of  Figs~\ref{fig:double_diff_p_mb} and~\ref{fig:double_diff_a_mb} are conflicting. 
Nonetheless, we can still comment on possible causes for the apparent discrepancy.

The simplest mechanisms that may increase the cross section in this region is a larger value for the axial form-factor. 
Calculations of Refs.~\cite{Meyer:2022mix, Simons:2022ltq} have shown that by using form-factors obtained in LQCD an increase of the cross section with up to $20\%$ is possible.
However, one cannot draw this conclusion without a full evaluation of the two-body current contribution, which can lead to a very similar enhancement.
Indeed, while similar calculations in the EDRMF are in closer agreement with the T2K data~\cite{Nikolakopoulos:2022qkq}, and inclusive electron scattering~\cite{Gonzalez-Jimenez19}, these do not include a realistic spectral function. The reduced occupancy of shell-model states would lead to an underprediction of the QE-peak in $(e,e')$ data.
However, the inclusion of spectroscopic factors for the shells is necessary for the description of the longitudinal response in inclusive electron scattering~\cite{Franco-Munoz:2022jcl} and $(e,e' p)$ data.
As such it is paramount to evaluate the contribution of two-body currents to one-nucleon knockout for neutrino interactions~\cite{Franco-Munoz:2022jcl, Lovato:2023khk}. 
The works of refs.~\cite{PhysRevC.51.2664, Franco-Munoz:2022jcl, Lovato:2023khk}, that use a similar approach as in this work, showed that the interference between one- and two-body currents leads to an increase of the transverse response. In combination with spectroscopic factors from $(e,e' p)$ experiments, this improves the description of the inclusive electromagnetic responses and cross section.
These findings are similar to what is found in the ab-initio calculations of Refs.~\cite{Lovato16, RevModPhys.87.1067, PhysRevC.65.024002}, in the sense that the inclusion of two-body electromagnetic currents can lead to an increase of the transverse vector-vector response by up to 20\%.
The GFMC calculations of Refs.~\cite{PhysRevX.10.031068, Lovato:2017cux}, showed that for the axial-axial contribution to electroweak responses, the longitudinal response is also modified.

\section{Summary and Conclusions}\label{conclusions}
We have generated fully differential scattering events for neutrino induced single-proton knockout from argon averaged over the MicroBooNE flux.
We include a realistic model for the nuclear spectral function, and perform calculations in both the relativistic distorted-wave impulse approximation (RDWIA) and plane wave approximation (RPWIA). The former uses the real energy-dependent potential of Refs.~\cite{Gonzalez-Jimenez19}, which is necessary to reproduce the inclusive cross section.
These events are used as input to several intranuclear cascade models (INCs). These are the INCs included in NEUT, \nuwro, \textsc{Achilles} and the Li\`ege INC (\incl).

As in Ref.~\cite{Nikolakopoulos:2022qkq}, we compare the output of these INCs with RDWIA calculations using the relativistic optical potential (ROP) of Ref.~\cite{Cooper93}. The latter approach is extensively used in the analysis of $(e,e'p)$ experiments, in particular for the recent extraction of argon and titanium spectral functions~\cite{PhysRevD.105.112002, PhysRevD.107.012005}.
We find reasonable agreement between the ROP calculations and the \incl results (with RPWIA inputs) and the \textsc{Achilles} results (with EDRMF inputs) for variables that probe the transverse kinematic imbalance.
Similar agreement as in \textsc{Achilles} is found with \nuwro, when the effect of SRC on the mean free path~\cite{Niewczas:2019fro} is not included.
For NEUT, the agreement is not as good due to a large transparency for low energy nucleons, as previously shown in Ref.~\cite{Nikolakopoulos:2022qkq}.
We don't find full agreement between any INC and the ROP calculations, which is made clear by considering the distribution of proton kinetic energies $T_p$.
The main differences found between the INCs are for low-energy nucleons. The treatment of short-range correlations in \nuwro, and the formation time effect included in \textsc{Achilles} lead to pronounced differences over the whole $T_p$ range.
For $T_p > 400~\mathrm{MeV}$, the INCs and the ROP agree on the shape of the $T_p$-dependence, and both the RPWIA and EDRMF inputs give similar results. The cross section obtained with the INCs is consistently around $25\%$ larger than the ROP result in this region however.
This difference points to a systematic inconsistency between the INC and the ROP approaches.

We assess the influence on the choice of spectral function for observables in MicroBooNE by considering several spectral functions for argon, and for nuclei in the same mass range (${}^{40}$Ca, and ${}^{56}$Fe).
We assess the sensitivity to these spectral functions in the PWIA. We find that MicroBooNE flux-averaged observables are insensitive to the large variations in missing energy distributions considered.
Variables that do not correlate lepton and hadron variables are found to be almost insensitive to the choice of spectral function altogether. This is in agreement with the findings of Ref.~\cite{Franco-Patino:2023msk}, which performed a systematic study using a parameterized spectral function. 
The distribution of $\delta P_T$ is sensitive to the missing momentum distributions. These are almost universal for the different nuclei considered, as would be expected for nuclei in this mass range.
We conclude that the effect of possible variations in a \emph{realistic} spectral function on the comparison to data are subdominant to the effects of final-state interactions.

We confront the calculations with MicroBooNE data, using both the RPWIA and EDRMF calculations as input to the INCs.
The EDRMF provides a reduction of the cross section compared to the RPWIA of around $10\%$. This reduction is quite consistent for all flux-averaged MicroBooNE observables considered.
The reduction comes mostly in the region of forward scattered muons.

The \nuwro, NEUT, and \textsc{Achilles} INCs give similar results in comparison to the MicroBooNE data.
The \incl is the outlier, and provides a stronger reduction at low $\delta P_T^{rec}$ than the other INCs, in addition to a large contribution in the high-$\delta P_T$ tail, even at small $\alpha_T$.

We point out that \incl predicts the production of a significant amount of nuclear clusters, most notably helium and deuteron. We find that $8\%$ of all events that pass the MicroBooNE selection cuts has at least one deuteron. This percentage rises to $18\%$ when no cuts on the hadron kinematics are considered.
We show that the production of deuterons with large momenta ($p > 500~\mathrm{MeV}$) is predicted. These should be experimentally measurable in a LArTPC.

All calculations generally underpredict the MicroBooNE data. This is expected, as two-nucleon knockout contributions from meson-exchange currents, and meson production are not included in the calculation.
These mechanisms should contribute mostly at large $\alpha_T$ and large $\delta P_T$. However, we find a significant lack of strength also in the low-$\delta P_T^{rec}$ region.
An increase at low-$\delta P_T$ can be provided by an increase of the axial coupling. However, the present results are expected to underpredict even inclusive electron scattering data, as the inclusion of the spectral function removes strength from the quasielastic peak. It was shown that the interference with two-body currents~\cite{Franco-Munoz:2022jcl, Lovato16}, can yield the required increase of the cross section in electron scattering.
It is hence important that this interference contribution is evaluated in predictions for neutrino interactions~\cite{Lovato:2023khk, VanCuyck17}.

 \begin{acknowledgments}
 We are indebted to J.M. Ud\'ias for innumerable valuable discussions on the shell model and optical potentials, and for making numerical codes for the distortion of the final nucleon available to us. We thank A. Ankowski for information on the spectral function; A. Furmanski for providing estimates for deuteron propagation in argon; N. Steinberg, A. Papadopoulou, V. Pandey, and R. Plestid for useful discussions.
 R.G.-J. is supported by Project No.~PID2021-127098NA-I00 funded by MCIN\slash AEI\slash 10.13039\slash 501100011033\slash FEDER, UE. 
 The work of J.I. and N.R. was supported by the U.S. Department of Energy, Office of Science, Office of Advanced Scientific Computing Research, Scientific Discovery through Advanced Computing (SciDAC-5) program, grant “NeuCol”.
 A.E. is supported by Laboratoire Leprince-Ringuet, Ecole polytechnique, IN2P3-CNRS, Palaiseau, France.
 K.N. acknowledges the support of the Fund for Scientific Research Flanders (FWO) and Ghent University Special Research Fund.
This work was supported by Fermi Research Alliance, LLC under Contract No. DE-AC02-07CH11359 with the U.S. Department of Energy, Office of Science, Office of High Energy Physics.
A. M. Kelly was supported in part by the U.S. Department of Energy, Office of Science, Office of Workforce Development for Teachers and Scientists (WDTS) under the Science Undergraduate Laboratory Internships Program (SULI).

 \end{acknowledgments}

\bibliographystyle{apsrev4-1.bst}
\bibliography{bibliography}

\onecolumngrid
\newpage

\appendix

\section{Hadron tensor}
\label{app:hadronT}
We use several approaches for the description of the hadron tensor in this work. These are described in this appendix.
The nuclear spectral function used in this work is made up of the contributions of several shells, each with specific angular momenta.
The hadron tensor for nucleon knockout out of a shell, labelled by the relativistic angular momentum and principal quantum numbers $\kappa$ and $n$ is given by
\begin{equation}
    \rho_{n,\kappa} (E_m) H^{\mu\nu}_{n,\kappa} = \frac{\rho_{n,\kappa}(E_m)}{2J+1} \sum_{m_j} \sum_{s_N} \left[ J^\mu_{n,\kappa} (Q,k_N, s_N, m_j) \right]^* J^{\nu}_{n,\kappa} (Q,k_N, s_N, m_j).
\end{equation}
Here $Q$ is the four-momentum transferred to the nucleus, and $k_N$ is the outgoing nucleon's four-momentum. The spin of the outgoing nucleon and the angular momentum projection of the bound-state single-particle wavefunction are $s_N$ and $m_j$ respectively.
\subsection{Relativistic distorted-wave impulse approximation (RDWIA)}
The four-current in the RDWIA is 
\begin{equation}
\label{eq:RDWIA}
    J^{\mu}_{n,\kappa} (Q,k_N, s_N, m_j) = \int \diff \mathbf{p}~ \overline{\psi}_{s_N}(k_N, \mathbf{p})~\mathcal{O}^\mu(Q)~\psi_\kappa^{m_j}(\mathbf{p} - \mathbf{q})
\end{equation}
The outgoing nucleon wavefunction $\psi_{s_N}(k_N, \mathbf{p}) = \frac{1}{(2\pi)^{3/2}}\int \diff \mathbf{r}~ e^{-i\mathbf{p}\cdot\mathbf{r}} \psi_{s_N}(k_N, \mathbf{r})$ is obtained in a partial wave expansion
\begin{equation}
    \label{eq:psi_scattering}
    \psi_{s_N}(k_N,\mathbf{r} ) = 4\pi \sqrt{\frac{E_N+M}{2M}} \sum_{\kappa, m_j, m_l} e^{i\delta_\kappa} i^{l} ( l~m_l ; \frac{1}{2}~m \vert j~m_j ) Y_{l, m_l}^* \left( \Omega_{N} \right) \Psi^{m_j}_{\kappa} \left( E_N, \mathbf{r} \right).
\end{equation}
The relativistic angular momentum enumerates states with definite spin and parity: $\lvert \kappa \rvert = j + 1/2$ and $l = j - 1/2 $ if $\kappa < 0$ and $l = j + 1/2$ when $\kappa > 0$. The phase shifts $\delta_\kappa$ are obtained by matching to the Dirac-Coulomb phase shift~\cite{Greiner}. 
The wavefunction $\Psi^{m_j}_{\kappa}(E,\mathbf{r})$ is of the form 
\begin{equation}
\label{eq:psi_kappa_r}
    \Psi_\kappa^{m_j}(E, \mathbf{r}) =  
\begin{pmatrix}
g_{\kappa}(E, \lvert \mathbf{r} \rvert) \Phi_{\kappa}^{m_j}(\Omega_r) \\
if_{\kappa}(E, \lvert \mathbf{r} \rvert) \Phi_{-\kappa}^{m_j}(\Omega_r)
\end{pmatrix},
\end{equation}
with the two-component spin-spherical harmonics
\begin{equation}
    \Phi_{\kappa}^m\left(\Omega \right) = \sum_{s} ( l, m - s ; \frac{1}{2}, s | \left. j \: m \right) Y_{l}^m\left(\Omega \right) \chi^{s}.
\end{equation}
The radial wavefunctions are obtained as a static scattering solution of the Dirac equation with energy-dependent vector and scalar potentials,
\begin{equation}
    i\slashed{\partial} \Psi_\kappa^{m_j}(E, \mathbf{r}) e^{iEt} = \left[ \left( M_N + S(E, \lvert \mathbf{r} \rvert ) \right) + \gamma^{0}V(E, \lvert \mathbf{r} \rvert ) \right] \Psi_\kappa^{m_j}(E, \mathbf{r}) e^{iEt}.
\end{equation}
The energy $E = \sqrt{\mathbf{p}_N^2 + M_N^2} = \omega - E_m - T_X + M_N$, is the asymptotic energy of the nucleon. 
In the following two subsections, the operator, and the bound-state wavefunction are discussed.

\subsection{Relativistic plane-wave impulse approximation (RPWIA)}
In the RPWIA, the final-state nucleon is described by a Dirac plane wave and Eq.~(\ref{eq:RDWIA}) reduces to
\begin{equation}
\label{eq:Jmu_RPWIA}
    J^{\mu}_{n,\kappa} (Q,k_N, s_N, m_j) = (2\pi)^{3/2}~\overline{u}(k_N, s_N) \mathcal{O}^\mu (Q) \psi_{\kappa}^{m_j}(\mathbf{p}_m = \mathbf{p}_N - \mathbf{q}).
\end{equation}
The bound-state single-particle wavefunction in momentum space is 
\begin{equation}
    \psi_\kappa^{m}(\mathbf{p}) = (-i)^l 
\begin{pmatrix}
g(p) \Phi_{\kappa}^m(\Omega_p) \\
if(p) \Phi_{-\kappa}^m(\Omega_p)
\end{pmatrix} = \frac{1}{(2\pi)^{3/2}} \int \mathrm{d} \mathbf{r} e^{-i\mathbf{p}\cdot\mathbf{r}} \Psi_{\kappa}^{m_j} (\mathbf{r}),
\end{equation}
with $\Psi_\kappa^{m_j}(\mathbf{r})$ a bound-state solution of the form of Eq.~(\ref{eq:psi_kappa_r}).
With this, it is straightforward to compute the four-current in the RPWIA. 
To make the connection to the PWIA we consider the hadron tensor obtained from Eq.~(\ref{eq:Jmu_RPWIA}),
\begin{equation}
    H^{\mu\nu}_{n,\kappa} = \frac{(2\pi)^3}{2J+1} \mathrm{Tr} \left\{ \frac{\slashed{k}_N + M_N}{2M_N} \mathcal{O}^\mu \left[\sum_{m_j} \psi_{\kappa}^{m_j}(\mathbf{p}_m) \overline{\psi}_\kappa^{m_j}(\mathbf{p}_m) \right] \gamma^0\mathcal{O}^\nu \gamma^0   \right\}.
\end{equation}
It is straightforward to show that~\cite{Bechler_1993},
\begin{equation}
    \sum_m \Phi^{m}_\kappa \left[ \Phi_\kappa^{m} \right]^\dagger = \frac{2J+1}{8\pi}\mathbf{I}, \quad \sum_m \Phi^{m}_\kappa \left[ \Phi_{-\kappa}^{m} \right]^\dagger = -\frac{2J+1}{8\pi} \frac{\sigma\cdot \mathbf{p}}{\lvert \mathbf{p} \rvert},
\end{equation}
by which the sum over $m_j$ can be performed. The results can be conveniently written as~\cite{GardnerPiekarewicz}
\begin{equation}
\label{eq:BS_prop}
    \sum_{m_j} \psi_{\kappa}^{m_j}(\mathbf{p}) \overline{\psi}_\kappa^{m_j}(\mathbf{p}) = \frac{2J+1}{4\pi} \left[\slashed{P}_\kappa(\mathbf{p}) + M_{\kappa}(\mathbf{p}) \right].
\end{equation}
Here $P_\kappa^\mu (\mathbf{p}) = (E_\kappa, \mathbf{p}_\kappa)$, with
\begin{equation}
\label{eq:BS_prop_components}
E_\kappa(\mathbf{p}) = f^2_\kappa(\lvert \mathbf{p} \rvert) + g^2_\kappa(\lvert \mathbf{p} \rvert ),\quad \mathbf{p}_\kappa(\mathbf{p}) = 2\frac{\mathbf{p}}{\lvert \mathbf{p} \rvert} f_\kappa(\lvert \mathbf{p} \rvert) g_\kappa(\lvert \mathbf{p} \rvert), \quad M_\kappa(\mathbf{p}) =  g^2_\kappa(\lvert \mathbf{p} \rvert) -  f^2_\kappa(\lvert \mathbf{p} \rvert).
\end{equation}
It is then straightforward to obtain analytical results in terms of the momentum-space wavefunctions.
The operator used in this work is
\begin{equation}
    \mathcal{O}^\mu = F_1 \gamma^\mu + i \frac{F_2}{2M_N} \sigma^{\mu\alpha} Q_{\alpha} + F_A \gamma^\mu \gamma^5 + F_P Q^\mu \gamma^5.
\end{equation}
With $F_1$ and $F_2$ the Dirac and Pauli form-factors from Kelly~\cite{Kelly04}. The axial form factor is parameterized as a dipole with $M_A=1~\mathrm{GeV}$, and $F_P = 4M_N^2 F_A (Q^2 + M_\pi^2)^{-1}$.
The hadron tensor can be separated in contributions from vector-vector axial-axial and vector-axial 
\begin{equation}
    H^{\mu\nu} = \frac{(2\pi)^
    3}{4\pi} \left[ H_{VV}^{\mu\nu} + H_{AA}^{\mu\nu} +i H^{\mu\nu}_{VA} \right].
\end{equation}
These are given by
\begin{align}
2M_N H^{\mu\nu}_{VV} &= 4 F_1^2 \left[k_N^\mu P_\kappa^\nu + P_\kappa^\mu k_N^\nu + g^{\mu\nu} \left( M M_\kappa - P_\kappa \cdot k_N \right) \right] \nonumber \\ &
+ 4F_1\frac{F_2}{M_N} \left[ g^{\mu\nu} \left( (M_\kappa k_N - M_N P_\kappa)\cdot Q \right) + \frac{ M_N (P_\kappa^\nu Q^\mu + P_\kappa^\mu Q^\nu) - M_\kappa (k_N^\nu Q^\mu + Q^\nu k_N^\mu)}{2} \right] \nonumber \\
&+ 4 \frac{F_2^{2}}{4 M_N^2} [  g^{\mu\nu} \left( P_\kappa \cdot k_N Q^2 -2 k_N\cdot Q P_\kappa\cdot Q + M_N M_\kappa Q^2 \right) + k_N \cdot Q \left(P_\kappa^\mu Q^\nu + P_\kappa^\nu Q^\mu \right)  \nonumber \\
&+ P_\kappa \cdot Q (k_N^\mu Q^\nu + k_N^\nu Q^\mu )  - Q^\mu Q^\nu (k_N \cdot P_\kappa + M_N M_\kappa ) - Q^2 ( k_N^\mu P_\kappa^\nu + k_N^\nu P_\kappa^\mu )  ] 
\label{eq:H_VV_RPWIA}
\end{align}
\begin{align}
2M_N H^{\mu\nu}_{AA} &=  4F_A^2\left( k_N^\mu P_\kappa^\nu + k_N^\nu P_\kappa^\mu  - g^{\mu\nu}(M_N M_\kappa + P_\kappa\cdot k_N )  \right)  \nonumber \\
&- 4F_P^2 Q^\mu Q^\nu \left(M_N M_\kappa - P_\kappa \cdot k_N  \right)  \nonumber \\
	&+ 4F_A F_P\left( Q^\mu(M_M P_\kappa^\nu - M_\kappa k_N) + Q^\nu ( M_N P_\kappa^\mu - M_\kappa k_N^\mu) \right) 
 \label{eq:H_AA_RPWIA}
\end{align}
\begin{align}
2M_N H^{\mu\nu}_{VA} = 4 \left[ 2 F_A F_1 \epsilon^{\mu\nu\rho\sigma} P_{\kappa,\rho} k_{N,\sigma}  -  2 F_A \frac{F_2}{2 M_N} \epsilon^{\mu\nu\beta\alpha}\left( M_N P_{\kappa,\beta} + M_\kappa k_{N,\beta} \right)Q_\alpha + F_P \frac{F_2}{4 M_N^2} [ Q^\mu \epsilon^{\nu\sigma\rho\alpha} - Q^\nu \epsilon^{\mu\sigma\rho\alpha} ] k_{N,\sigma} Q_{\rho} P_{\kappa, \alpha} \right].
\label{eq:H_VA_RPWIA}
\end{align}
\subsection{Plane-wave impulse approximation (PWIA)}
The difference of the expressions obtained in the RPWIA, with respect to commonly used expressions in the PWIA~\cite{PhysRevC.102.064626, VOrden2019}, stem from the fact that the plane-wave particle spinors do not form a complete basis for the expansion of the bound-state wavefunctions. 
Indeed, the full basis involves anti-particle spinors as well, as explored in much detail in Refs.~\cite{Caballero:1997gc}. 
In the PWIA, the anti-particle components are either projected out explicitly, or it is generally posited that the bound-state projector (Eq.~\ref{eq:BS_prop}) is proportional to a free-nucleon projector. 
The hadron tensor is then of the form
\begin{equation}
    H^{\mu\nu}_{PWIA} = (2\pi)^3 \alpha^2(p_m) Tr \left\{ (\slashed{k}_N + M_N) \mathcal{O}^\mu \left(\slashed{\overline{k}}_m + M_N \right) \gamma^0 \mathcal{O}^\nu  \gamma^0 \right\},
\end{equation}
with $\overline{k}_m = (\sqrt{\mathbf{p}_m^2 + M_N^2}, \mathbf{p}_m)$.

To derive such a factorized expression, we can construct a bound-state without negative energy component as in Ref.~\cite{Caballero:1997gc}.
When the relation between upper and lower components that follows from the positive energy Dirac equation holds,
\begin{equation}
    \psi_d = \frac{\sigma\cdot\mathbf{p}}{\overline{E}+M} \psi_u,
\end{equation}
where $\overline{E} = \sqrt{\mathbf{p}^2 + M_N^2}$.
This leads to $f(p) = \sqrt{(\overline{E}-M)/(\overline{E}+M)} g(p)$, and Eq.~(\ref{eq:BS_prop_components}) reduces to
\begin{equation}
    E_\kappa = \frac{\overline{E}}{\overline{E}+M} g^2(p), \quad M_\kappa = \frac{M}{\overline{E}+M} g^2(p), \quad \mathbf{p}_\kappa = \frac{\mathbf{p}}{E+M} g^2(p)
\end{equation}
which gives $4\pi \alpha^2(p_m) = \frac{g^2(p_m)}{\overline{E} + M}$.
Comparing this to the momentum distribution
\begin{equation}
    n(p) = \frac{g^2(p) + f^2(p)}{4\pi} = \frac{2\overline{E}}{\overline{E}+M} \frac{g^2(p)}{4\pi}, 
\end{equation}
we can write $\alpha^2(p_m) = n(p_m)/2\overline{E}$. 
We can thus write the semi-exclusive cross section in the factorized form
\begin{equation}
    \frac{d\sigma(E_\nu)}{dp_\mu d\Omega_\mu d\Omega_p dp_N} = \frac{G_F^2 \cos^2\theta_c}{(2\pi)^2} \frac{p_\mu^2 p_N^2}{E_\nu E_\mu} \frac{M_N^2}{E_N \overline{E}}  L_{\mu\nu} h_{s.n.}^{\mu\nu} S(E_m, p_m),
\end{equation}
with here $S(E_m, p_m) = \sum_\kappa N_\kappa \rho_\kappa(E_m) n(p_m)$.
The single-nucleon hadron tensor can be directly obtained from the RPWIA expressions by substitution of $\slashed{P}_\kappa + M_\kappa \rightarrow (\slashed{\overline{k}}_m + M_N)/(2M_N)$.
One obtains the same result for the hadron tensor as in Refs.~\cite{PhysRevC.102.064626}. 

For the results in this work, we have followed Ref.~\cite{Benhar08} and substituted $Q \rightarrow \overline{Q} \equiv k_N - \overline{k}_m$ in the operator, but not in the form factors. 
In this case the hadron tensor simplifies and can be written as
\begin{equation}
    h_{s.n.}^{\mu\nu} = \frac{1}{M_N^2} \left\{ W_1 M_N^2 g^{\mu\nu} + W_2 \overline{k}_m^\mu \overline{k}_m^\nu + i W_3 \epsilon^{\mu\nu\alpha\beta} \overline{k}_{m,\alpha} \overline{Q}_\beta + W_4 \overline{Q}^\mu \overline{Q}^\nu + W_5 (\overline{k}_m^\mu \overline{Q}^\nu + \overline{k}_m^\nu \overline{Q}^\mu)  \right\}
\end{equation}
with
\begin{align}
    W_1 &= - \tau \left( F_1 + F_2\right)^2 - F_A^2(\tau + 1), \\
    W_2 &= F_1^2 + \tau F_2^2 + G_A^2, \\
    W_3 &= G_A\left( F_1 + F_2\right), \\
    W_4 &= \frac{\tau -1}{4} F_2^2 - \frac{F_1 F_2}{2} - F_P G_A M_N + \tau M_N^2  F_P^2, \\
    W_5 &= \frac{W_2}{2},
\end{align}
where $\tau = -\overline{Q}\cdot\overline{Q}/(4M_N^2)$.

\section{Alternative options in cascade models}\label{app:alternate}
We show here additional comparisons between the INC and optical potential calculations that use a different treatment of correlations.

\begin{figure*}
\includegraphics[width=0.98\textwidth]{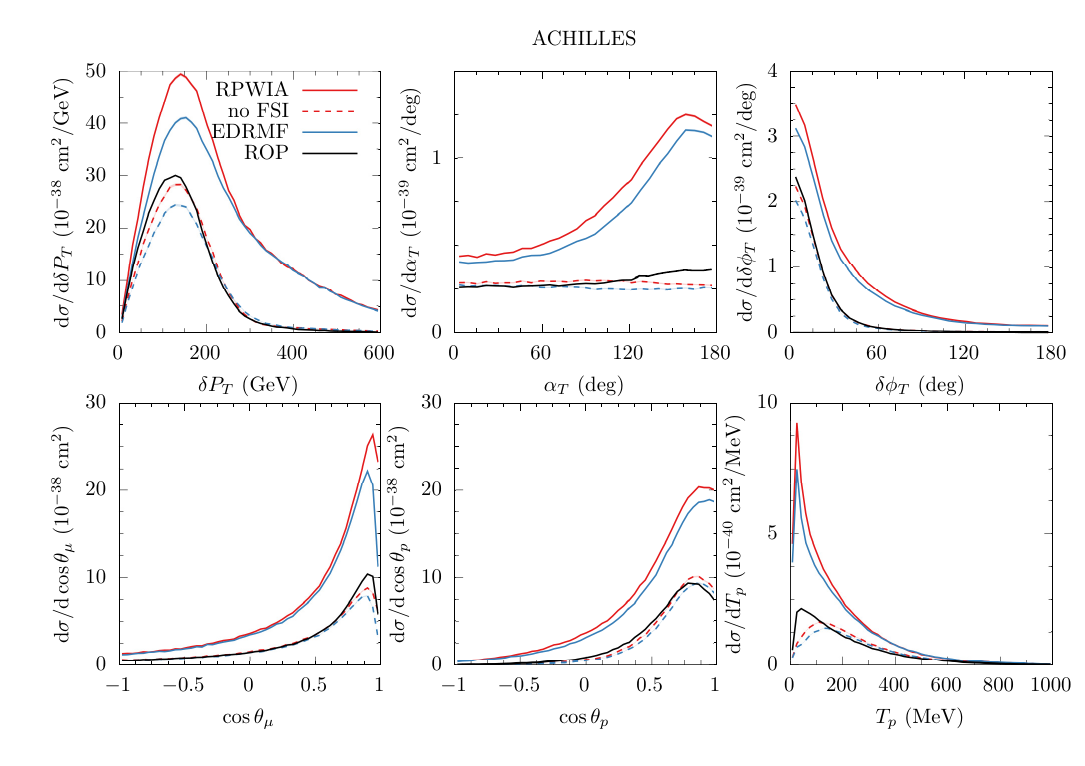} 
\caption{Same as in Fig.~\ref{fig:NEUT_compROP}, but for \textsc{Achilles} where the effect of the formation time in the initial interaction is not included.}
\label{fig:INC_ROP_ACHILLES_noFZ}
\end{figure*}

In Fig.~\ref{fig:INC_ROP_ACHILLES_noFZ}, we show the comparison between the \textsc{Achilles} INC and the ROP calculations. The difference with respect to Fig.~\ref{fig:ACHILLES_compROP} is that a formation time effect is included in the latter. In the present comparison, this effect is not included.
This leads to a decrease of the nuclear  transparency. 

\begin{figure*}
\includegraphics[width=0.98\textwidth]{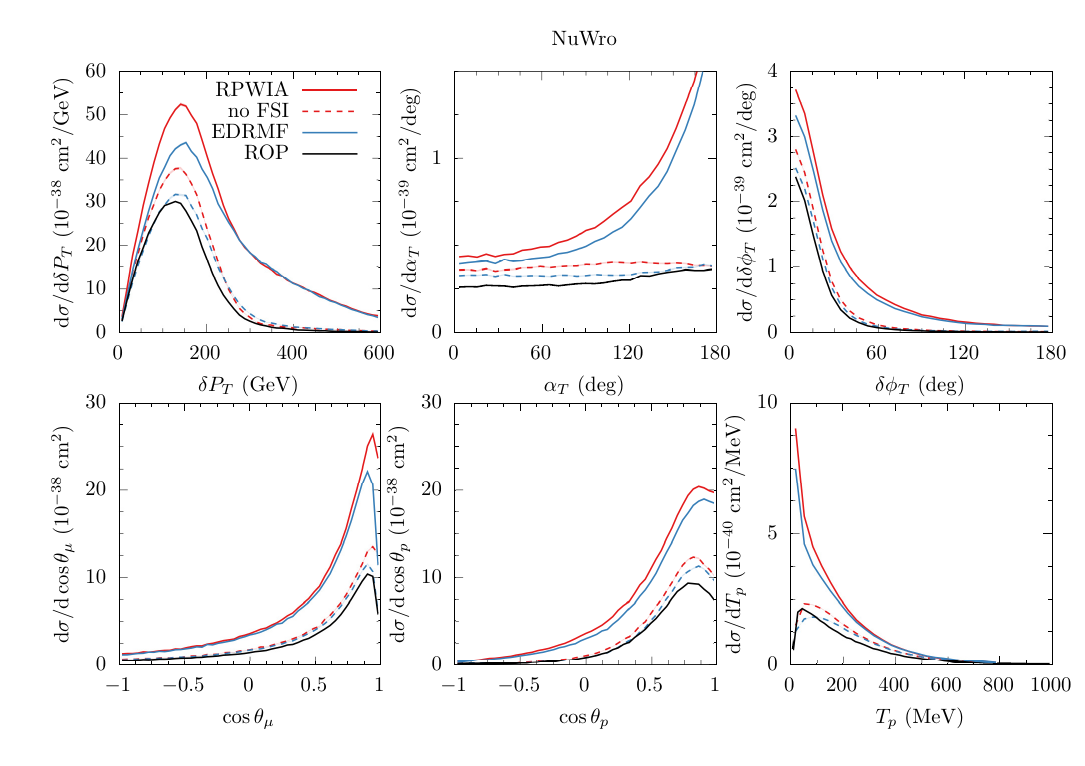} 
\caption{Same as in Fig.~\ref{fig:NEUT_compROP}, but for \nuwro where the effect of SRC on the two-nucleon density~\cite{Niewczas:2019fro} is not included.}
\label{fig:INC_ROP_NUWRO_noSRC}
\end{figure*}

In Fig.~\ref{fig:INC_ROP_NUWRO_noSRC}, we show the comparison of \nuwro and the ROP, this time without including the effect of nucleon-nucleon correlations on the mean-free path of the struck nucleon.

\end{document}